\documentclass[12pt]{article}
\usepackage[english]{babel}
\usepackage[latin1]{inputenc}

\usepackage[intlimits]{amsmath}
\usepackage{amssymb}

\usepackage{url}
\usepackage{graphicx}
\usepackage{slashed}
\usepackage{color}

\numberwithin{equation}{section}

\newcommand{\GeV}{\mbox{GeV}}
\newcommand{\TeV}{\mbox{TeV}}
\newcommand{\cm}{\mbox{cm}}

\def\lsim{\mathrel{\raise.3ex\hbox{$<$\kern-.75em\lower1ex\hbox{$\sim$}}}}
\def\gsim{\mathrel{\raise.3ex\hbox{$>$\kern-.75em\lower1ex\hbox{$\sim$}}}}

\begin{document}


\title{
{\normalsize
TUM-HEP 808/11\hfill\mbox{}\\}
\vspace{1.5cm} 
\bf Antiproton constraints on dark matter annihilations
from internal electroweak bremsstrahlung \\[8mm]}

\author{Mathias Garny, Alejandro Ibarra, Stefan Vogl\\[2mm]
{\normalsize\it Physik-Department T30d, Technische Universit\"at M\"unchen,}\\[-0.05cm]
{\it\normalsize James-Franck-Stra\ss{}e, 85748 Garching, Germany}
}

\maketitle

\begin{abstract}
\noindent
If the dark matter particle is a Majorana fermion, 
annihilations into two fermions and one gauge boson
could have, for some choices of the parameters of the model, 
a non-negligible cross-section. Using a toy
model of leptophilic dark matter, we calculate
the constraints on the annihilation cross-section
into two electrons and one weak gauge boson from the
PAMELA measurements of the cosmic antiproton-to-proton
flux ratio. Furthermore, we calculate the maximal
astrophysical boost factor allowed in the Milky Way 
under the assumption that the leptophilic dark matter
particle is the dominant component of dark matter in
our Universe. These constraints constitute very conservative
estimates on the boost factor for more realistic models
where the dark matter particle also couples to quarks
and weak gauge bosons, such as the lightest neutralino
which we also analyze for some concrete benchmark points.
The limits on the astrophysical boost factors presented here 
could be used to evaluate the prospects to detect a gamma-ray
signal from dark matter annihilations at 
currently operating IACTs as well as in the projected CTA.
\end{abstract}

\newpage

\section{Introduction}

The indirect search of dark matter has entered in the last few years
into a new era of precision measurements.
The Fermi Large Area Telescope (LAT) has measured
with unprecedented accuracy the flux of cosmic electrons with energies
between 7 GeV and 1 TeV~\cite{Ackermann:2010ij} and the flux of
cosmic gamma-rays between approximately 20 MeV and 
300 GeV~\cite{Atwood:2009ez}. Besides, the satellite-borne experiment 
PAMELA has measured the cosmic antiproton flux and the antiproton-to-proton
fraction between 60 MeV and 180 GeV~\cite{Adriani:2010rc}, the 
electron flux between 1 and 625 GeV~\cite{Adriani:2011xv}, and the
positron to electron fraction between 1.5 and 100 GeV~\cite{Adriani:2008zr}.

The quality of the data collected by these 
experiments opens excellent opportunities
to search for the annihilation or the decay of dark matter particles
with masses in the electroweak range. Unfortunately, this search
is hindered by the existence of large, and poorly known, astrophysical
backgrounds which complicates in general the identification of a dark 
matter component in cosmic rays against the astrophysical backgrounds.

This generic difficulty can be circumvented in some situations.
The most notable one is when the dark matter particle annihilates or
decays producing monoenergetic gamma-rays, which produces 
a very peculiar signal in the energy spectrum which cannot be 
mimicked by any known astrophysical source. Conversely, the non-observation 
of a gamma-ray line in the Fermi-LAT data sets very stringent constraints 
on the dark matter annihilation cross-section or decay rate into
 monoenergetic gamma rays~\cite{Abdo:2010nc,Vertongen:2011mu}.

On the other hand, when the dark matter is a Majorana particle, higher
order processes such as $\chi\chi\rightarrow f^+f^- \gamma$ can
become important for some choices of the parameters of the model, 
despite being suppressed by the extra coupling
and by the phase space factor. In this case, gamma rays
can be produced in the final state radiation, i.e. radiated off a nearly
on-shell $f^\pm$ particle, or in the (virtual) internal bremsstrahlung,
radiated off the internal bosonic line or from an off-shell $f^\pm$ particle~\cite{Bringmann:2007nk,Bergstrom:1989jr,Flores:1989ru}. Whereas
in the former process the energy spectrum of photons is rather featureless, and
hence difficult to disentangle from the also featureless
spectrum produced by astrophysical sources, the latter
displays a very prominent bump which could be easily discriminated
from the background. Several papers have recently analyzed the constraints
on the annihilation cross-section into gamma-rays via internal
bremsstrahlung from the Fermi-LAT measurements of the diffuse gamma-ray
background~\cite{Barger:2009xe} or the dwarf galaxy
Segue 1~\cite{Scott:2009jn}, from the H.E.S.S. observations of
the globular clusters NGC 6388 and M 15~\cite{HESS:2011hh}
and the dwarf galaxies Sculptor and Carina~\cite{HESS:2010zzt}, as well
as from  the MAGIC observations of the dwarf galaxy
Willman 1~\cite{Aliu:2008ny}.
Furthermore, the prospects to observe this gamma-ray signature at
currently operating Imaging Air Cherenkov Telescopes (IACTs)
was evaluated in~\cite{Bringmann:2008kj,Viana:2011tq}
and at the projected Cerenkov Telescope Array (CTA)
in~\cite{Bringmann:2008kj,Viana:2011tq,Cannoni:2010my}.

Due to the $SU(2)_L\times U(1)_Y$ invariance of the Standard Model,
the annihilation process $\chi\chi\rightarrow f^+f^- \gamma$
is necessarily accompanied by the process
$\chi\chi\rightarrow f^+f^- Z$, which leads to the production
of antiprotons through the fragmentation of the $Z$ boson. Additional
antiprotons can arise through the fragmentation of the fermions, provided
they carry baryon number. Given the very stringent limits on the existence
of an exotic component in the antiproton-to-proton fraction, 
it is important to analyze the constraints on the parameters of the
model from the PAMELA measurements, in order to 
evaluate the prospects to observe the associated gamma-ray signal at present and
future experiments.

In this paper we will consider a toy dark matter model, where the dark matter
particle is a Majorana fermion, singlet under the Standard Model gauge
group, which couples to the electron lepton doublet and to a scalar doublet
via a Yukawa coupling. Despite the dark matter particle being ``leptophilic''
the annihilation will necessarily produce antiprotons via 
electroweak bremsstrahlung processes. In Section \ref{sec:pp} we will present the
details of the model and we will calculate the cross-sections for the 
$2\rightarrow 3$ processes involving two fermions in the final state, as well
as a photon, a $Z$ boson or a $W$ boson for different values of the 
parameters. In Section \ref{sec:obs}, we will calculate the limits on the
cross-section for the processes involving weak gauge bosons from the PAMELA
measurements on the antiproton-to-proton fraction. Finally, in Section
\ref{sec:neutralino} we investigate implications for realistic models, in
particular for several MSSM benchmark models that have been considered in
the past in connection with internal bremsstrahlung. Various analytical
expressions for $2\rightarrow 3$ processes can be found in
Appendix\,\ref{app:sigma}.

\section{Annihilations into two fermions and one gauge boson}\label{sec:pp}

We consider a toy model, consisting in extending the particle content of the Standard Model by a Majorana fermion, $\chi=\chi^c$, which we assume to constitute the dark matter of the Universe, and a scalar field, $\eta$. Their quantum numbers under the gauge group $SU(3)_c\times SU(2)_L\times U(1)_Y$ are:
\begin{align}
\chi\equiv (1,1,0)\;, ~~~~~
\eta=\begin{pmatrix} \eta^+ \\ \eta^0 \end{pmatrix}\equiv (1,2,1/2)\;.
\end{align}
Namely, $\chi$ is a singlet under the gauge group while $\eta$ has the same gauge quantum numbers as the Standard Model Higgs. We further impose that the field $\eta$ carries electron lepton number $L_e=-1$ in order to forbid its couplings to the quarks as well as to the second and third lepton generations\footnote{A similar model, with lepton number softly broken, has been considered in \cite{Ma:2000cc} in connection to neutrino masses.}.
With this particle content, the Lagrangian reads:
\begin{align}
{\cal L}={\cal L}_{\rm SM}+{\cal L}_{\chi}+{\cal L}_\eta+
{\cal L}^{\rm fermion}_{\rm int}+{\cal L}^{\rm scalar}_{\rm int}\;.
\end{align}
Here, ${\cal L}_{\rm SM}$ is the Standard Model Lagrangian which includes
a potential for the Higgs doublet $\Phi$, $V=m_1^2 \Phi^\dagger \Phi
+\frac{1}{2}\lambda_1 (\Phi^\dagger \Phi)^2$. On the other hand 
${\cal L}_\chi$ and ${\cal L}_\eta$ are the parts of the Lagrangian
involving just the Majorana fermion $\chi$ and the scalar particle $\eta$,
respectively, and which are given by
\begin{align}
\begin{split}
{\cal L}_\chi&=\frac12 \bar \chi^c i\slashed{\partial} \chi
-\frac{1}{2}m_\chi \bar \chi^c\chi\;, \\
{\cal L}_\eta&=(D_\mu \eta)^\dagger  (D^\mu \eta)-m_2^2 \eta^\dagger\eta-
\frac{1}{2}\lambda_2(\eta^\dagger \eta)^2\;,
\end{split}
\end{align}
where $D_\mu$ denotes the covariant derivative.
Lastly, ${\cal L}^{\rm fermi}_{\rm int}$ and ${\cal L}^{\rm scalar}_{\rm int}$
denote the fermionic and scalar interaction terms of the new
particles to the electron doublet and to the Higgs doublet:
\begin{align}
\begin{split}
{\cal L}^{\rm fermion}_{\rm int}&= f \bar \chi(L_e i\sigma_2 \eta)+{\rm h.c.}=
f \bar \chi (\nu_{eL} \eta^0-e_L \eta^+)+{\rm h.c.}\;,  \\
{\cal L}^{\rm scalar}_{\rm int}&= -\lambda_3(\Phi^\dagger \Phi)(\eta^\dagger \eta)
-\lambda_4(\Phi^\dagger \eta)(\eta^\dagger \Phi)\;.
\end{split}
\end{align}

We will assume that $m_1^2<0$ and $m_2^2>0$, therefore the minimization
of the potential leads to $\langle \phi^0\rangle=\sqrt{-m_1^2/\lambda_1}$,
 $\langle \eta^0\rangle=0$. The mass spectrum of the scalars can be
straightforwardly calculated, the result being:
\begin{align}
\begin{split}
m^2_{\phi^0}&=2 \lambda_1 v_{EW}^2 \;,\\
m^2_{\eta^0}&=m_2^2+(\lambda_3+\lambda_4)v_{EW}^2\;,   \\
m^2_{\eta^\pm}&=m_2^2+\lambda_3 v_{EW}^2\;.
\end{split}
\end{align}
Note that $\eta^0$ is, like the charged component, a complex field that carries lepton number and contains the CP-even and CP-odd neutral components of the doublet $\eta$.
In order to guarantee the stability of the dark matter particle, we require that $m_\chi< m_{\eta^0,\eta^\pm}$. For dark matter masses in the range 100 GeV-  TeV, as we will assume here, this condition is automatically satisfied if $m_2\gg M_Z$. 

The quartic couplings in the scalar potential are subject to constraints from demanding the absence of unbounded from below directions in the potential. These conditions were derived in \cite{Deshpande:1977rw}:
\begin{align}
\begin{split}
&\lambda_1>0, ~~~\lambda_2>0,~~~\lambda_3>-(\lambda_1 \lambda_2)^{1/2}\;,  \\
& \lambda_3+\lambda_4>-(\lambda_1 \lambda_2)^{1/2}\;.
\end{split}
\end{align}
These four conditions can be simultaneously fulfilled for both signs of the coupling $\lambda_4$. Therefore both orderings $m^2_{\eta^\pm}<m^2_{\eta^0}$ and $m^2_{\eta^\pm}>m^2_{\eta^0}$ are theoretically viable, which as we will see has implications in determining which is the dominant annihilation channel of the dark matter particle.

The Majorana dark matter particle $\chi$ can be produced thermally in the early universe, and the correct abundance could be achieved, for a given dark matter mass, by adjusting the coupling constant $f$~\cite{Cao:2009yy} (neglecting coannihilations, and assuming a freeze-out temperature $T_{fo}=m_{DM}/20$ and $g_*(T_{fo})=106.75$, see e.g. \cite{Drees:2009bi}):
\begin{align}\label{eq:omegaDM}
\Omega_{DM} h^2\simeq 0.11 \left(\frac{0.35}{f}\right)^4 \left(\frac{m_{DM}}{100\GeV}\right)^2 \left[ \frac{1+m_{\eta^\pm}^4/m_{DM}^4}{(1+m_{\eta^\pm}^2/m_{DM}^2)^4} + \frac{1+m_{\eta^0}^4/m_{DM}^4}{(1+m_{\eta^0}^2/m_{DM}^2)^4} \right]^{-1}\;.
\end{align}
In principle, this would fix also the absolute strength of any signals relevant for dark matter detection. However, since we are here mostly interested in a comparative analysis of different indirect dark matter signals, and since the production could be influenced by additional heavier degrees of freedom not included in the toy model or by non-thermal effects, we will not a priori restrict the value of $f$ in the following (unless stated otherwise) but rather assume that the particle $\chi$ constitutes the dark matter in the present universe with the correct abundance.

The observational prospects for detecting energetic particles produced in annihilations of dark matter particles are determined by the velocity-weighted annihilation cross-section.
It can be expanded into s- and p-wave contributions,
\begin{equation}
 \sigma v = a + b v^2 + \mathcal{O}(v^4) \;.
\end{equation}
For typical relative dark matter velocities $v\lesssim 10^{-3}c$ within the Milky Way halo \cite{Ascasibar:2003mm}, the p-wave contribution $bv^2$ is usually strongly suppressed.

The lowest-order annihilation channels are the two-to-two processes $\chi\chi\to e\bar e$ and $\chi\chi\to \nu\bar\nu$, which proceed by exchanging an $\eta^\pm$-particle or an $\eta^0$-particle, respectively. Because of the Majorana nature of $\chi$, these processes can proceed via the t-  and u-channel. The interference of these diagrams leads to the well-known helicity-suppression of the s-wave contribution to the  two-to-two annihilation cross-section, $a\propto m_e^2/m_{DM}^2$. In the limit $m_e=0$, the cross-section for the
channel $\chi\chi\to e\bar e$ is given by~\cite{Cao:2009yy}
\begin{equation}
 a \approx 0, \qquad b = \frac{f^4}{48\pi m_{DM}^2}\frac{1+m_{\eta^\pm}^4/m_{DM}^4}{(1+m_{\eta^\pm}^2/m_{DM}^2)^4}\;.
\end{equation}
For $\chi\chi\to \nu\bar\nu$ one has to replace $m_{\eta^\pm}\mapsto m_{\eta^0}$.
Since the s-wave contribution is helicity suppressed, and the p-wave contribution is suppressed by $v^2$, the annihilations $\chi\chi\to e\bar e$ and $\chi\chi\to \nu\bar\nu$ are not efficient in the Milky Way today.

Additional annihilation channels arise at higher orders in perturbation theory, e.g. the two-to-three process $\chi\chi\to \gamma e\bar e$. Due to additional couplings and phase-space, one would naively expect a suppression factor of order $\alpha_{em}/\pi\sim 2\cdot 10^{-3}$ compared to the two-to-two process. As is well-known, the emission of a soft or collinear photon off the final electron (or positron) is logarithmically enhanced $\sim \alpha_{em}/\pi[\ln(m_e^2/(4m_{DM}^2))]^2$, and leads to a model-independent spectrum characteristic for final state radiation (e.g.~\cite{Birkedal:2005ep}).

More importantly, the emission of a photon can also lift the helicity suppression. The reason is that the three-body final state $\gamma e\bar e$ can simultaneously accommodate a left-handed electron, a right-handed positron, and possess zero angular momentum. Therefore, the s-wave contribution is not helicity suppressed. Consequently, the annihilation channel $\chi\chi\to \gamma e\bar e$ can even dominate over the lowest-order one, because the absence of helicity suppression more than compensates the suppression factor $\alpha_{em}/\pi$~\cite{Bringmann:2007nk}.

Note that this feature is an important difference to the soft or collinearly emitted bremsstrahlung photons, which yield only a comparably small contribution. The copious emission of hard and non-collinear photons has been called `virtual internal bremsstrahlung', because the emission can be viewed as a part of the hard process. We will refer to this process simply as internal bremsstrahlung (IB) in the following. Typically, the spectrum of the produced photons is peaked at high energies, of the order of the dark matter mass. Therefore, these photons lead to a characteristic gamma-ray signal that can be potentially observed by present and future gamma-ray telescopes~\cite{Bringmann:2008kj,Viana:2011tq,Cannoni:2010my}.

In addition to the two-to-three process $\chi\chi\to \gamma e\bar e$ producing a photon, $SU(2)_L\times U(1)_Y$-invariance inevitably leads to the presence of the two-to-three process $\chi\chi\to Z e\bar e$, provided the channel is kinematically allowed, i.e. $m_{DM}>M_Z/2$. As for the photon, one can discriminate
between ordinary final state radiation and (virtual) internal bremsstrahlung. In the former case, the $Z$-boson is radiated off an on-shell electron. Although this process cannot lift helicity suppression,
it is subject to a logarithmic enhancement and can be important for heavy dark matter particles, $m_{DM}\gg\TeV$ \cite{Berezinsky:2002hq,Bell:2008ey,Kachelriess:2009zy,Ciafaloni:2010ti}.

On the other hand, internal electroweak bremsstrahlung describes the emission of a $Z$-boson off the internal line, or from an off-shell electron. This process can lift the helicity suppression, like the photon. It is the main purpose of this work to study these electroweak IB processes, for which one can expect a similar enhancement as for the electromagnetic IB process $\chi\chi\to \gamma e\bar e$. The further decay and fragmentation of the $Z$-boson produces in particular protons and antiprotons, in equal amounts. In addition, the decay products lead to additional photon, electron and positron as well as neutrino production.

In view of the stringent observational constraints on the cosmic flux of antiprotons, it is important to study the process $\chi\chi\to Z e\bar e$, in particular when considering model parameters that lead to a strong gamma-ray signal from electromagnetic IB. Within the toy model considered here, also the electroweak processes $\chi\chi\to Z \nu\bar \nu$, $\chi\chi\to W \bar e\nu$, and $\chi\chi\to W e\bar\nu$ contribute to the production of antiprotons.

\begin{figure}
 \includegraphics{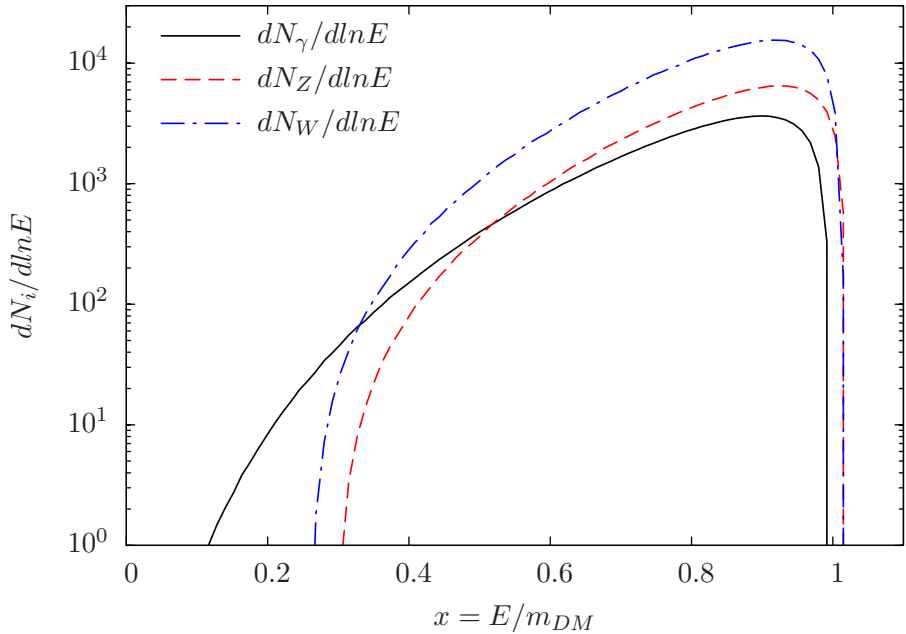}
 \caption{\label{fig:spectrum1} Spectrum of photons as well as $Z$- and $W$-bosons produced by internal bremsstrahlung in dark matter annihilations for $m_{DM}=300\GeV$ and $m_{\eta^\pm}=m_{\eta^0}=330\GeV$.}
\end{figure}

The energy distribution of $W$-bosons produced by IB in dark matter annihilations can be represented by the dimensionless differential cross-section
\begin{equation}
 \frac{dN_W}{d\ln E} = \frac{1}{\sigma v(\chi\chi\to e\bar e)} \left( \frac{vd\sigma(\chi\chi\to W\bar e\nu)}{d\ln E} + \frac{vd\sigma(\chi\chi\to W e\bar \nu)}{d\ln E}\right) \,,
\end{equation}
which is defined analogously to the corresponding quantity $dN_\gamma/d\ln E=\sigma v(\chi\chi\to e\bar e)^{-1}\,vd\sigma(\chi\chi\to \gamma e\bar e)/d\ln E$ for electromagnetic IB. The normalization to the two-to-two cross-section $\sigma v(\chi\chi\to e\bar e)$ is chosen as in \cite{Bringmann:2007nk} for electromagnetic IB for convenience. Similarly, one can also define $dN_Z/d\ln E$.

Our analytic results for the s-wave contribution to the differential cross-section can be found in appendix~\ref{app:sigma}. They are applicable for arbitrary mass spectrum, except for neglecting the electron mass $m_e=0$, in the relevant range $M_W/2<m_{DM}<m_{\eta^\pm},m_{\eta^0}$, in particular for masses $m_{\eta^i}$ nearly degenerate with $m_{DM}$ and also for $m_{\eta^\pm}\not=m_{\eta^0}$. The numerical results shown in the following include also velocity-suppressed contributions\footnote{We have used CALCHEP \cite{Pukhov:1999gg, Pukhov:2004ca} for parts of the analytical as well as for numerical computations.}.

The $W$, $Z$, and photon spectra from IB are shown in Fig.\,\ref{fig:spectrum1}. The massive gauge bosons are produced with a hard spectrum that is peaked near the maximal energy, similar to electromagnetic IB. However, the maximal energy is slightly higher, $E_{max}=m_{DM}+M_{Z(W)}^2/(4m_{DM})$. Another difference is the lower cut-off at the gauge boson mass. Since the shape of the spectra near the peak are nevertheless very similar, we will use the total cross-sections in order to illustrate the relative importance of the various processes in the following (we use the full spectra in our numerical analysis, however).

The relative strength of electromagnetic IB, $\chi\chi\to \gamma e\bar e$, and the electroweak IB process $\chi\chi\to Z e\bar e$, for dark matter masses far above the threshold $m_{DM}\gg M_Z/2$, is simply related by the relative couplings of (left-handed) electrons to photons and to $Z$-bosons, respectively,
\begin{equation}
  \sigma v(\chi\chi\to Z e\bar e) \ : \ \sigma v(\chi\chi\to \gamma e\bar e) = \cot^2(2\theta_W) = 0.41 \qquad\mbox{for }m_{DM}\gg M_Z/2\;.
\end{equation}
This is because both processes are mediated by the same particle, namely the charged component $\eta^\pm$ of the doublet. For the other electroweak IB processes, also the neutral component $\eta^0$ of the doublet appears in the relevant Feynman diagrams. Therefore, their branching ratios depend on the mass $m_{\eta^0}$, or, more specifically, on the mass splitting $m_{\eta^0}^2-m_{\eta^\pm}^2=\lambda_4 v_{EW}^2$.

\subsubsection*{The case $m_{\eta^0}=m_{\eta^\pm}$}

Let us first assume that the components of the doublet are degenerate, $m_{\eta^0}=m_{\eta^\pm}$, i.e. $\lambda_4=0$ in the scalar potential\footnote{This case has also been recently studied in Refs. \cite{Ciafaloni:2011sa,Bell:2011if}.}. Since both mediating particles are degenerate, all branching ratios are simply determined by the relative coupling constants  for $m_{DM}\gg M_Z/2$,
\begin{equation}
\begin{array}{cccccccr}
  \sigma v(\chi\chi\to Z \nu\bar \nu) & : & \sigma v(\chi\chi\to \gamma e\bar e)
  & = & \frac{1}{\sin^{2}(2\theta_W)} & = & 1.41\\
  \sigma v(\chi\chi\to W e\nu) & : & \sigma v(\chi\chi\to \gamma e\bar e)
  & = & \frac{1}{\sin^{2}(\theta_W)} & = & 4.32 &
\end{array}\ \mbox{for }m_{DM}\gg \frac{M_Z}{2}, m_{\eta^0}=m_{\eta^\pm}\;.
\end{equation}
Here $\sigma v(\chi\chi\to W e\nu)\equiv \sigma v(\chi\chi\to W^- \bar e\nu) + \sigma v(\chi\chi\to W^+ e\bar \nu)$.

\begin{figure}
\hspace*{-2cm}
\begin{tabular}{ll}
 \includegraphics[width=0.6\textwidth]{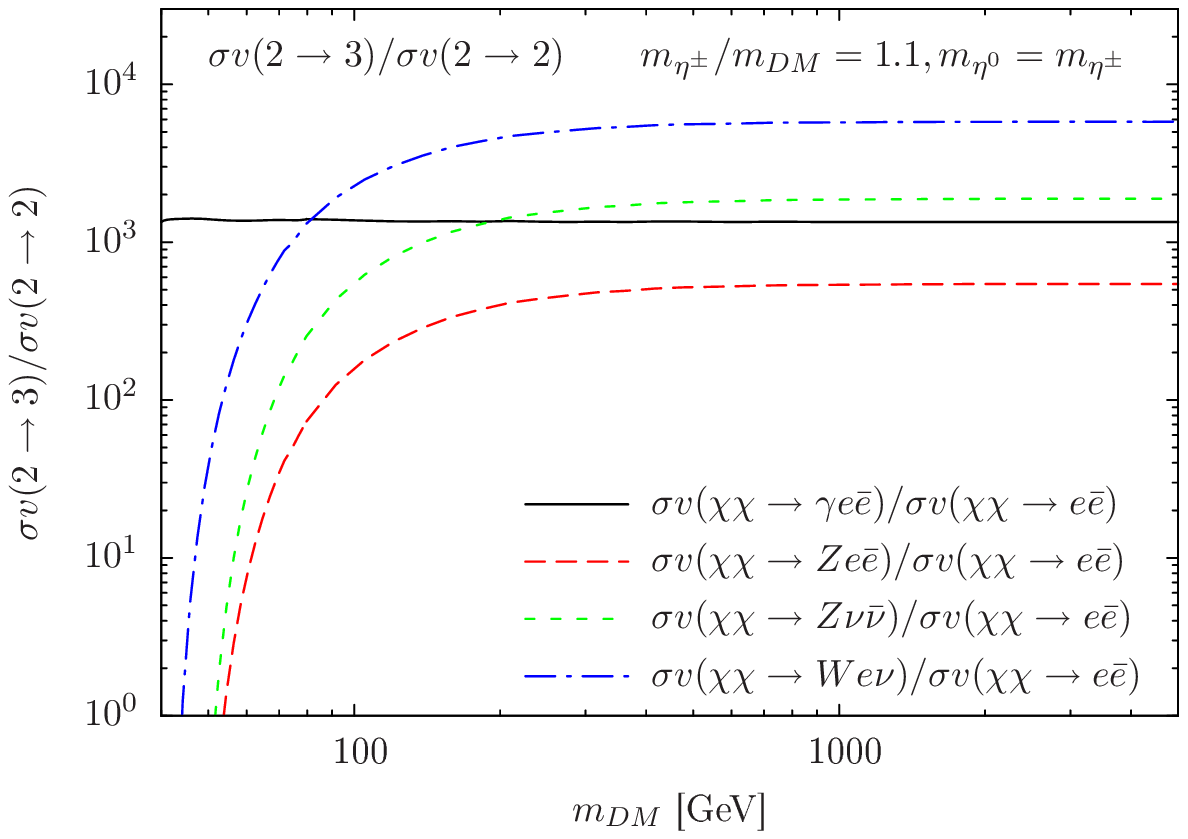}
 &\hspace*{-1.6cm} \includegraphics[width=0.6\textwidth]{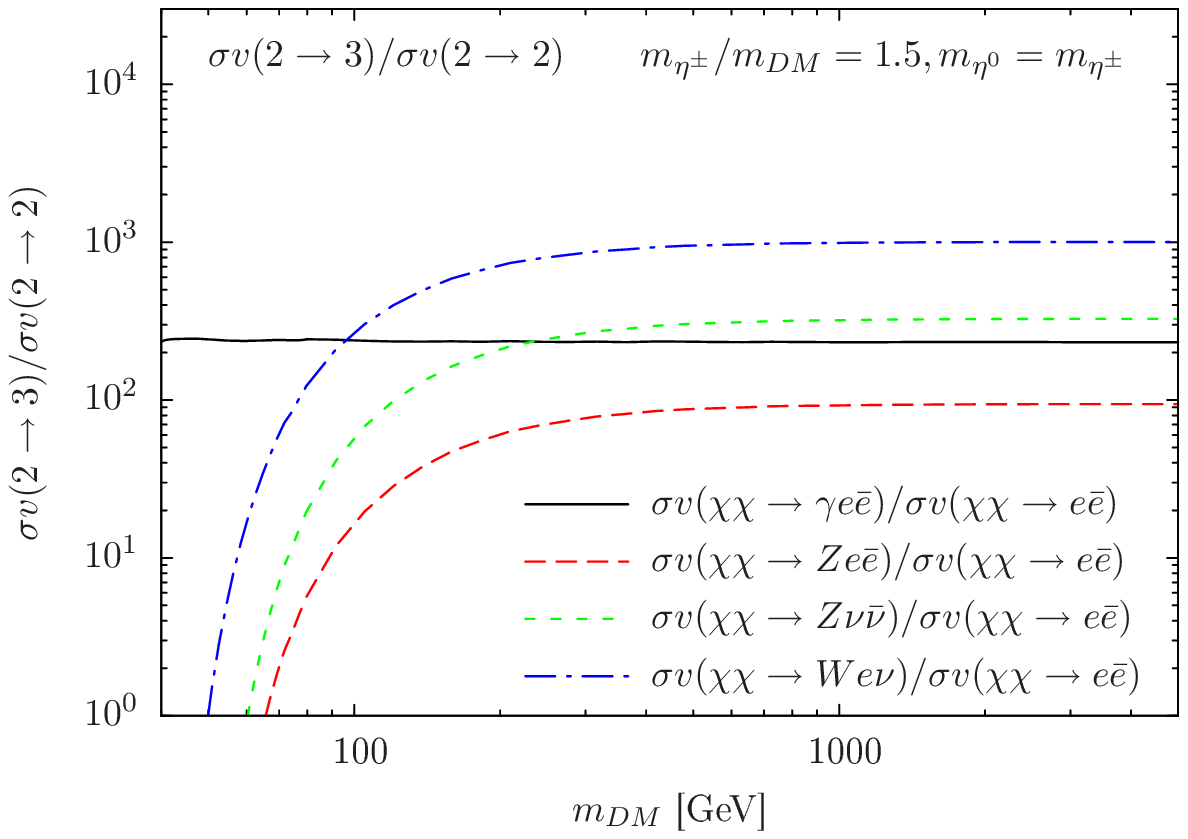}
\end{tabular}
 \caption{\label{fig:crossSection1} Ratio of three-body and two-body annihilation cross-sections, for electromagnetic IB, $\sigma v(\chi\chi\to \gamma e\bar e)/\sigma v(\chi\chi\to e\bar e)$, and for the electroweak IB channels $\chi\chi\to Ze\bar e, \chi\chi\to Z\nu\bar\nu$ and $\chi\chi\to We\nu$. The latter denotes the sum of $W^- \bar e\nu$ and $W^+ e\bar \nu$. The ratio of the scalar doublet mass to the dark matter mass is fixed to $m_{\eta^\pm}/m_{DM}=1.1$ (left) and $1.5$ (right), and here we assume that both components of the doublet have identical mass, $m_{\eta^0}=m_{\eta^\pm}$. For the relative dark matter velocity we use $v=10^{-3}c$.}
\end{figure}

The full result for the ratio of the two-to-three cross-sections to the two-to-two cross-section $\sigma v(\chi\chi\to e\bar e)$ is shown in Fig.\,\ref{fig:crossSection1} for $m_{\eta^0}=m_{\eta^\pm}$, as a function of the dark matter mass. Clearly, the two-to-three processes dominate over the two-to-two annihilation channel, i.e. all ratios are much larger than one. Furthermore, one can observe that the branching ratios obey the simple relations given above for dark matter masses $m_{DM}\gtrsim 200$GeV. For dark matter masses in the range $45 \GeV < m_{DM}\lesssim 200\GeV$, the branching ratios for electroweak IB become kinematically suppressed, but are still sizeable for $m_{DM}\gtrsim 100\GeV$. Since the energies we are interested in, i.e. at which IACTs are sensitive to observe a gamma-ray signal, typically lie above $\sim 100\GeV$, we generically expect an $\mathcal{O}(1)$ branching fraction into electroweak gauge bosons in addition to the gamma signal from IB.

\begin{figure}
 \includegraphics{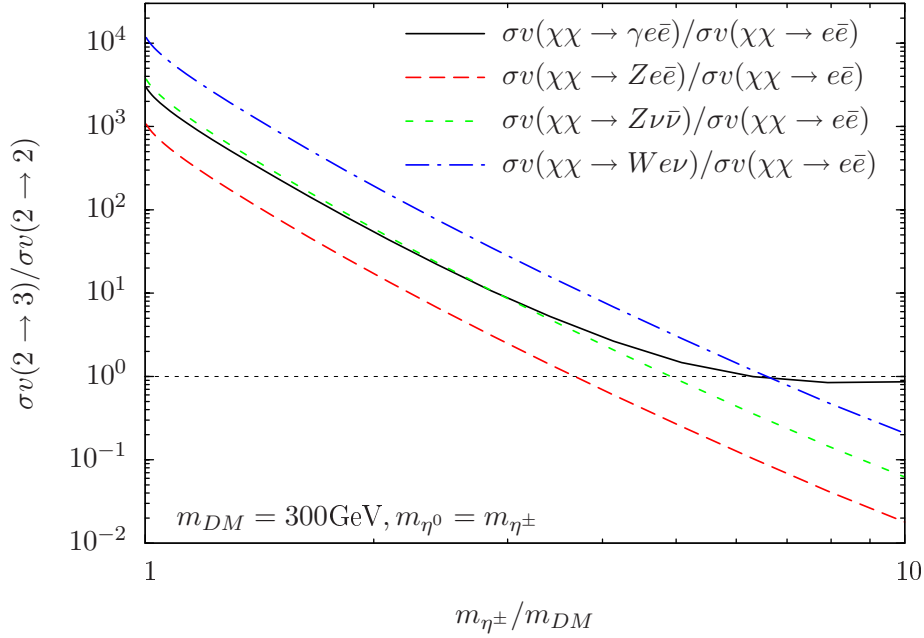}
 \caption{\label{fig:crossSection3} Ratio of three-body and two-body annihilation cross-sections for electromagnetic and electroweak IB, as a function of the doublet mass $m_{\eta^\pm}$. The dark matter mass is $m_{DM}=300$GeV, and $m_{\eta^0}=m_{\eta^\pm}$. Similarly to electromagnetic IB, the strongest enhancement of electroweak IB occurs when the dark matter mass $m_{DM}$ and the mass $m_{\eta^\pm}$ of the mediating particle are nearly degenerate. The IB annihilation channels dominate for $m_{\eta^\pm}\lesssim 5 m_{DM}$, i.e. when the masses are of comparable order of magnitude. For a strongly hierarchical spectrum, $m_{\eta^\pm}\gtrsim 5 m_{DM}$, the two-to-two channel as well as collinear electromagnetic final state radiation dominate.  For the relative dark matter velocity we use $v=10^{-3}c$.}
\end{figure}

In Fig.\,\ref{fig:crossSection3}, the ratio of two-to-three and two-to-two cross-sections are shown as a function of the mass $m_{\eta^\pm}$ of the particles ${\eta^\pm}$ and $\eta^0$ that are exchanged in the annihilation process, assuming $m_{\eta^\pm}=m_{\eta^0}$. Similar to electromagnetic IB, the electroweak IB is strongly enhanced when $m_{\eta^\pm}$ is nearly degenerate with $m_{DM}$. The IB processes dominate over the two-to-two channel as long as both masses are of the same order of magnitude, $1<m_{\eta^\pm}/m_{DM}\lesssim 5$. In the limit $m_{\eta^\pm}/m_{DM}\gg 1$, we find that the s-wave electroweak IB cross-sections scale like $\sigma v(2\to 3)|_{s-wave} \propto (m_{DM}/m_{\eta^\pm})^{8}$. This is the same scaling compared to electromagnetic IB. Since the two-to-two cross-section scales like $\sigma v(2\to 2) \propto (m_{DM}/m_{\eta^\pm})^{4}$, all the IB processes become suppressed when the mediating particles are very heavy, roughly $m_{\eta^\pm}/m_{DM}\gtrsim 5$. In that case the conventional final state radiation of soft and collinear photons would be the dominant production mechanism, which scales also with $(m_{DM}/m_{\eta^\pm})^{4}$, but cannot lift the helicity suppression. However, there is a large and generic region of parameter space where electroweak and electromagnetic IB are the dominating annihilation channels.

\subsubsection*{The case $m_{\eta^0}\not=m_{\eta^\pm}$}

Due to electroweak symmetry breaking, one expects in general that the neutral and charged components of the $SU(2)_L$ doublet $\eta$ acquire different masses. The mass difference is controlled by the order parameter $v_{EW}=174\GeV$ of electroweak symmetry breaking. In particular, within the toy model one finds that $m_{\eta^0}^2-m_{\eta^\pm}^2=\lambda_4 v_{EW}^2$, where $\lambda_4$ is a coupling in the scalar potential that is generally non-zero. Also, for example within the CMSSM, the components of the slepton doublet are non-degenerate at the electroweak scale. Therefore, it is important to discuss this case. For the toy model considered here, this case has not been discussed in the literature in connection with IB to our knowledge.

Clearly, the cross-sections for $\chi\chi\to \gamma e\bar e$ and $\chi\chi\to Z e\bar e$ are not affected by a mass-splitting, since they only involve the $\eta^\pm$ particle. Furthermore, the channel $\chi\chi\to Z \nu\bar\nu$ just involves the $\eta^0$ particle, and therefore (neglecting lepton masses)
\begin{equation}
\sigma v(\chi\chi\to Z \nu\bar\nu)={\textstyle \frac{1}{\cos^{2}(2\theta_W)}} \, \sigma v(\chi\chi\to Z e\bar e)|_{m_{\eta^\pm} \mapsto m_{\eta^0}} \approx 1.41 \left(\frac{m_{\eta^\pm}}{m_{\eta^0}}\right)^8 \sigma v(\chi\chi\to \gamma e\bar e)\,,
\end{equation}
where the latter estimate is valid for $m_{\eta^i} \gg m_{DM} \gg M_Z/2$. This is precisely the behaviour expected from simple scaling arguments. In particular, it results in an additional enhancement of electroweak IB when $m_{\eta^0} < m_{\eta^\pm}$. Since we are primarily interested in the case of a potentially strong gamma signal, we will assume that $m_{\eta^0} > m_{\eta^\pm}$, i.e. $\lambda_4>0$, in the following.

The most interesting process is the channel $\chi\chi\to We\nu$. The first reason is that this is the electroweak IB process which yields the largest contribution within the toy model, and is thus the crucial one for antiproton production. The second reason is related to the fact that both the neutral and charged components $\eta^0$ and $\eta^\pm$ simultaneously appear in the corresponding Feynman diagrams. In particular, the vertex $\eta^0\eta^\pm W^\mp$ features a coupling to both the transversally polarized as well as longitudinally polarized $W$-bosons, $W_T$ and $W_L$. The latter is related to the coupling to the Goldstone bosons $G^\pm$ that give mass to the $W$-bosons, according to the equivalence theorem. This coupling is given by
\begin{equation}
 \mathcal{L} \supset -\frac{g}{\sqrt{2}}\frac{m_{\eta^\pm}^2-m_{\eta^0}^2}{M_W}\eta^0\eta^+G^- + \mbox{h.c.} \;.
\end{equation}
Thus, the coupling is non-zero only in the presence of a mass splitting. This means that in the case $m_{\eta^0}=m_{\eta^\pm}$ discussed above only transversely polarized $W$-bosons are produced, $\chi\chi\to W_Te\nu$. When taking a mass-splitting into account, a new channel opens up, namely $\chi\chi\to W_Le\nu$. This feature represents a genuine difference compared to the IB of photons. Note that the upper vertex actually contributes in Feynman gauge, while it is implicitly taken into account in unitary gauge. We have checked that our results obtained in both gauges agree with each other.

\begin{figure}
\hspace*{-2cm}
\begin{tabular}{ll}
 \includegraphics[width=0.61\textwidth]{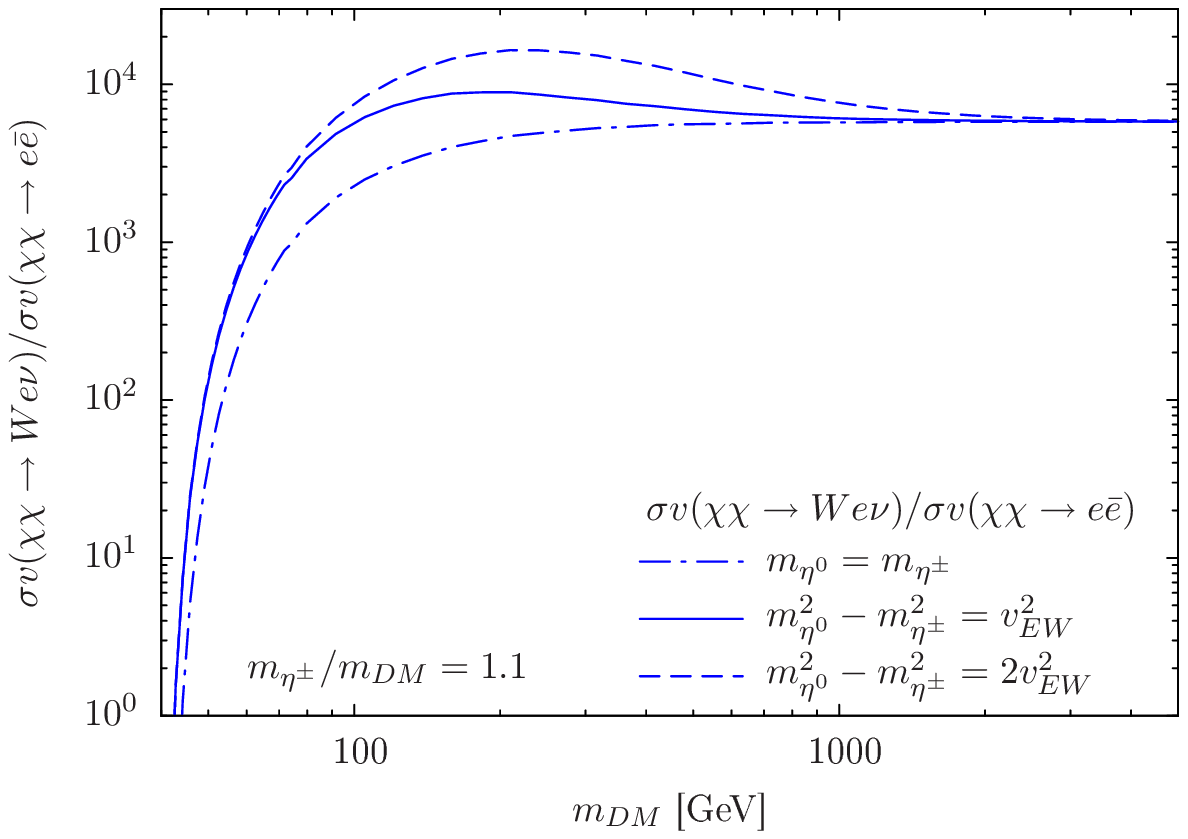}
 &\hspace*{-2cm} \includegraphics[width=0.61\textwidth]{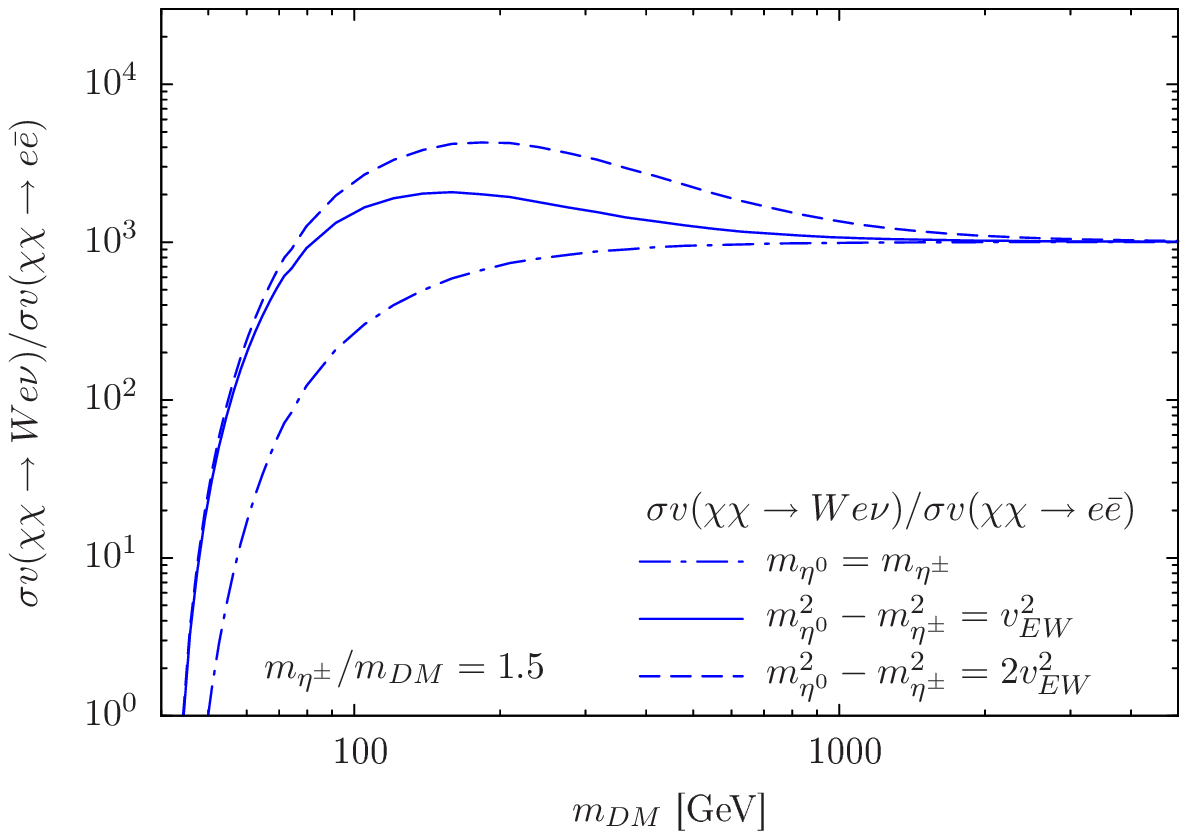}
\end{tabular}
 \caption{\label{fig:crossSection2} Ratio of three-body and two-body annihilation cross-sections $\sigma v(\chi\chi\to W e\nu)/\sigma v(\chi\chi\to e\bar e)$ for the electroweak IB channels into W-bosons ($W e\nu\equiv W^- \bar e\nu + W^+ e\bar \nu$).  The cross-section $\sigma v(\chi\chi\to W e\nu)$ depends sensitively on the mass-splitting $m_{\eta^0}^2 - m_{\eta^\pm}^2$ of the components of the doublet. For the previously considered case where $m_{\eta^0}=m_{\eta^\pm}$ (dot-dashed line), only transversely polarized W-bosons can be produced. For comparison, we show the cross-section when taking a mass-splitting into account (solid and dashed lines). In that case also longitudinally polarized W-bosons can be produced. In particular, the choice $m_{\eta^0}^2 - m_{\eta^\pm}^2=v_{EW}^2$ with $v_{EW}=174$GeV characterizes a typical splitting expected from electroweak symmetry breaking. The other parameters are chosen as in fig.\,\ref{fig:crossSection1}.}
\end{figure}

The annihilation $\chi\chi\to W_Le\nu$ is important for two reasons: first, we find that for a generic mass splitting $m_{\eta^0}^2-m_{\eta^\pm}^2=\lambda_4v_{EW}^2$ with $\lambda_4\sim\mathcal{O}(1)$, the annihilation into longitudinal $W$-bosons significantly enhances the total cross-section. This can be seen in Fig.\,\ref{fig:crossSection2}, where the case with mass splitting is compared to the degenerate case. The enhancement is important for dark matter masses up to $m_{DM}\sim 1-2\TeV$. A simple analytic estimate is possible for $m_{\eta^i} \gg m_{DM} \gg M_W/2$,
\begin{eqnarray}
\sigma v(\chi\chi\to We\nu) & \approx & \frac{1}{\sin^{2}(\theta_W)} \left(\frac{2m_{\eta^\pm}^2}{m_{\eta^0}^2+m_{\eta^\pm}^2}\right)^4 \left[ 1 + \frac{5}{8}\frac{(m_{\eta^0}^2-m_{\eta^\pm}^2)^2}{ M_W^2m_{DM}^2 }  \right] \sigma v(\chi\chi\to \gamma e\bar e)\nonumber\\
&\approx& 4.32 \left(\frac{2m_{\eta^\pm}^2}{m_{\eta^0}^2+m_{\eta^\pm}^2}\right)^4 \left[ 1 + \lambda_4^2\left(\frac{300\GeV}{m_{DM}}\right)^2  \right] \sigma v(\chi\chi\to \gamma e\bar e)\;,
\end{eqnarray}
where the second term in the square bracket corresponds to longitudinal $W$-bosons. More accurate expressions can be found in appendix\,\ref{app:sigma}.

\begin{figure}
 \includegraphics{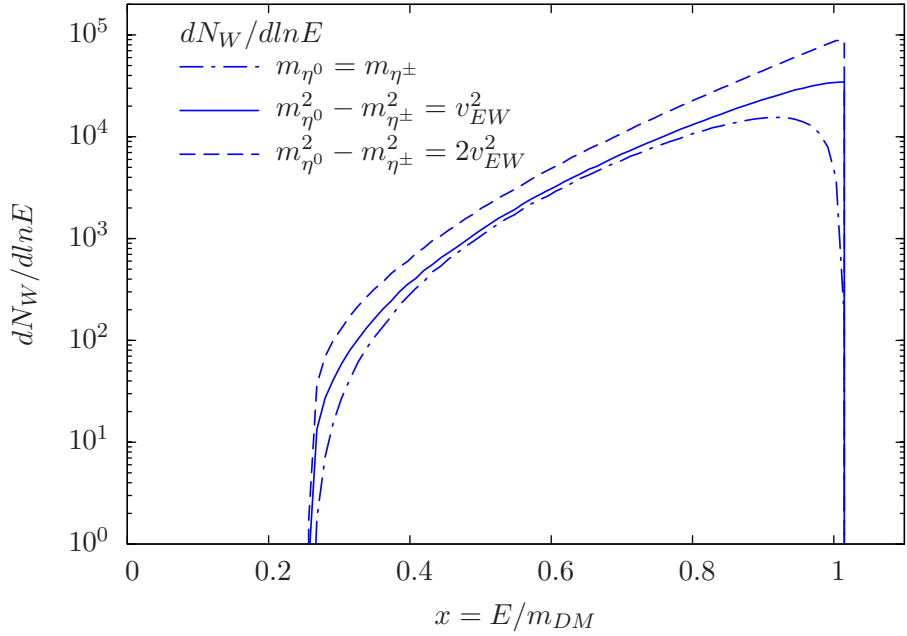}
 \caption{\label{fig:spectrum2} Spectrum of $W$-bosons produced by IB in dark matter annihilations for $m_{DM}=300\GeV$, $m_{\eta^\pm}=330\GeV$ and different choices of $m_{\eta^0}$. For $m_{\eta^0}\not= m_{\eta^\pm}$, also longitudinal $W$-bosons are produced, which can have the maximal energy $E_{max}=m_{DM}+M_W^2/4m_{DM}$ and therefore lead to a harder spectrum.}
\end{figure}

Second, also the spectrum of the produced longitudinal $W_L$ differs from the spectrum of the transverse $W_T$ considered before. The reason is that for $\chi\chi\to W_Le\bar\nu$, it is possible that the $e$ and $\bar\nu$ are emitted in one direction, while the $W_L$ is emitted in the opposite direction. This configuration supplies the $W_L$ with the maximal possible energy $E_{max}=m_{DM}+M_W^2/4m_{DM}$. For a transversal $W_T$, this configuration is forbidden by angular momentum conservation (up to a tiny helicity-suppressed contribution). Therefore, the spectrum of $W$-bosons obtained when taking a mass splitting into account is significantly harder than the spectrum obtained in the case $m_{\eta^0}=m_{\eta^\pm}$. This behaviour can be clearly seen in Fig.\,\ref{fig:spectrum2}, and is also manifest in the full analytic result presented in appendix\,\ref{app:sigma}.

In Fig.\,\ref{fig:crossSection4}, the scaling of the cross-section $\sigma v(\chi\chi\to We\nu)$ with $m_{\eta^\pm}$ is shown, when keeping the mass-squared difference $m_{\eta^0}^2-m_{\eta^\pm}^2=\lambda_4v_{EW}^2$ and the dark matter mass fixed. As before, the s-wave contribution scales with $(m_{DM}/m_{\eta^\pm})^8$. We find that the enhancement of the cross-section due to emission of longitudinal $W_L$ persists for large $m_{\eta^\pm}\gg m_{DM}$, even when keeping $m_{\eta^0}^2-m_{\eta^\pm}^2$ constant, as appropriate.

\begin{figure}
 \includegraphics{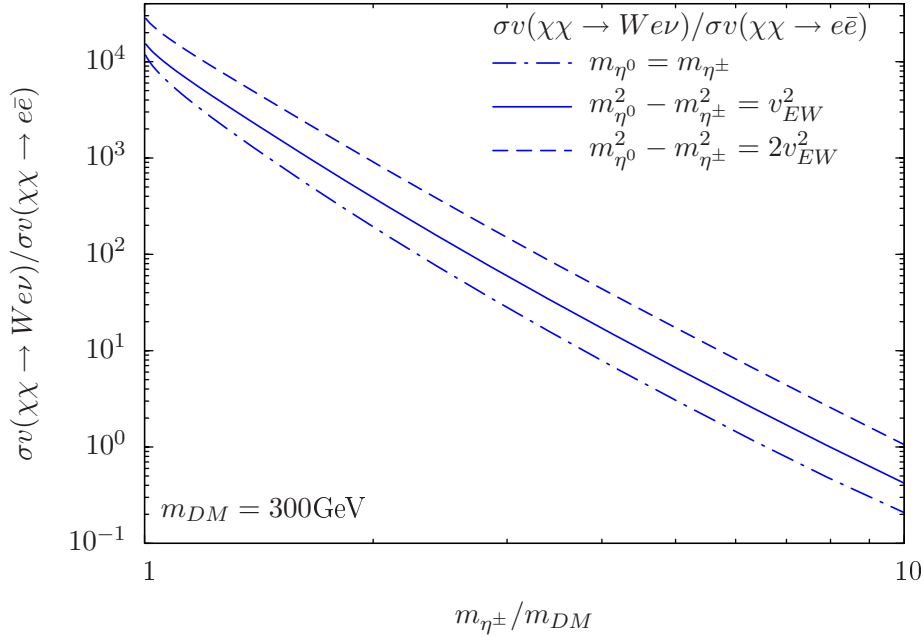}
 \caption{\label{fig:crossSection4} Ratio of three-body and two-body annihilation cross-sections $\sigma v(\chi\chi\to W e\nu)/\sigma v(\chi\chi\to e\bar e)$ for the electroweak IB channel into W-bosons, for different choices of the mass-splitting $m_{\eta^0}^2 - m_{\eta^\pm}^2$, as a function of the charged scalar mass $m_{\eta^\pm}$. The other parameters are chosen as in Fig.\,\ref{fig:crossSection3}.}
\end{figure}

\section{Antiproton flux  and observational constraints }\label{sec:obs}

We consider a distribution of dark matter particles in the Milky
Way given by $\rho(\vec r)$, where $\vec r$ denotes
the position of the dark matter particle with respect to the center
of our Galaxy.
In order to estimate the uncertainty in the predictions stemming
from our ignorance of the actual dark matter distribution in 
our Galaxy, we will calculate the predictions for three
different halo profiles. Concretely we will assume for simplicity
a spherically symmetric distribution with a radial dependence
given by the Isothermal profile
\begin{align}
\rho(r)=\frac{\rho_s}{1+(r/r_s)^2}\;,
\end{align}
the Navarro-Frenk-White (NFW) profile \cite{Navarro:1995iw,Navarro:1996gj}
\begin{align}
\rho(r)=\rho_s \frac{1}{r/r_s(1+r/r_s)^2}\;,
\end{align}
and the Einasto profile \cite{Navarro:2003ew,Graham:2006ae}
\begin{align}
\rho(r)=\rho_s \exp\left\{ 
-\frac{2}{\alpha}\left[\left(\frac{r}{r_s}\right)^\alpha-1
\right]\right\}\;.
\end{align}
In these expressions, the scale radius is $r_s=4.38,~24.42$ and 
$28.44\,{\rm kpc}$, respectively, while $\alpha=0.17$ for the Einasto profile. 
Besides, the parameter $\rho_s$ is adjusted in order
to yield a local dark matter density $\rho(r_\odot)=0.39 \,{\rm GeV}/{\rm cm}^3$~\cite{Catena:2009mf} with $r_\odot=8.5\,{\rm kpc}$ being the distance of 
the Sun to the Galactic center and is given, respectively, by 
$\rho_s=1.86,~0.25$ and $0.044\,{\rm GeV}/{\rm cm}^3$.

\begin{figure}
\hspace*{-1.6cm}
\begin{tabular}{ll}
 \includegraphics[width=0.59\textwidth]{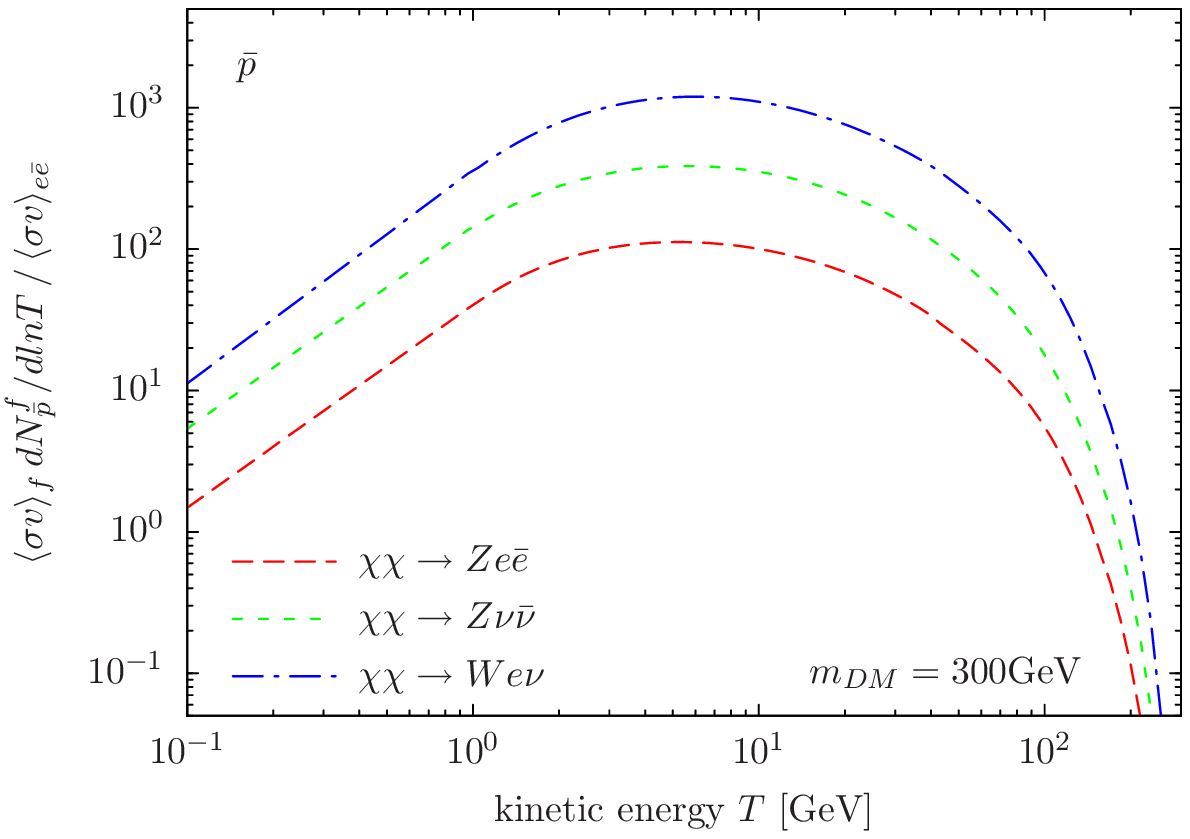}
 &\hspace*{-1.7cm} \includegraphics[width=0.59\textwidth]{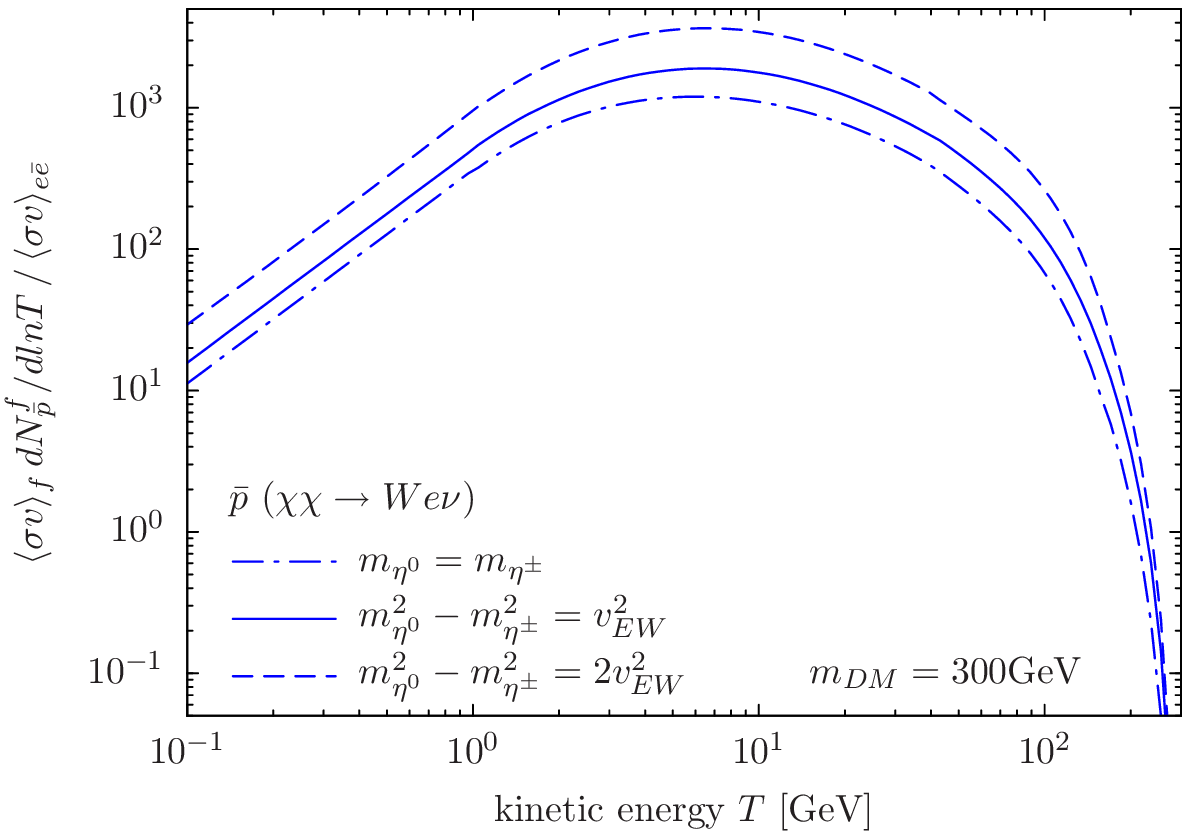}
\end{tabular}
 \caption{\label{fig:pbarspectrum} Spectrum of antiprotons (before propagation) resulting from the decay and fragmentation of the $W$- and $Z$-bosons produced in electroweak IB processes. Parameters are $m_{DM}=300\GeV$, $m_{\eta^\pm}=330\GeV$. The left plot shows the individual contributions for $m_{\eta^0}=m_{\eta^\pm}$, and the right one the $W$ contribution for different choices of the mass-splitting $m_{\eta^0}^2-m_{\eta^\pm}^2$.}
\end{figure}

Two dark matter particles at the position $\vec r$ can annihilate 
producing antiprotons at a rate per unit of kinetic energy and volume
given by:
\begin{align}
Q(T,\vec r)=\frac{1}{2}\frac{\rho^2(\vec r)}{m^2_\chi}
\sum_f \langle \sigma v\rangle_f \frac{dN^f_{\bar p}}{dT}\;,
\label{eq:source}
\end{align}
where $\langle \sigma v\rangle_f$ is the thermally averaged cross-section
multiplied by the velocity in the annihilation channel $f$
and $dN^f_{\bar p}/dT$ is the energy spectrum of antiprotons produced in 
that channel per unit of kinetic energy.

The weighted antiproton spectrum $\langle \sigma v\rangle_f \, dN^f_{\bar p}/d\ln\,T \, / \, \langle\sigma v\rangle_{\chi\chi\to e\bar e}$, normalized to the two-to-two cross-section is shown in Fig.\,\ref{fig:pbarspectrum}. The left part shows the spectrum resulting from the various electroweak IB bremsstrahlung processes and the right part the spectrum resulting from $W$-production for various values of the scalar mass-splitting. We have obtained the spectrum using the event generator PYTHIA 8.1 \cite{Sjostrand:2007gs} interfaced with CALCHEP \cite{Pukhov:1999gg, Pukhov:2004ca}.

After being produced at the position $\vec r$, antiprotons propagate
through the Milky Way in a complicated way before reaching the Earth.
Following \cite{ACR}, we will describe antiproton propagation by means of
a stationary two-zone diffusion model with cylindrical boundary conditions.
Under this approximation, the number density of antiprotons
per unit kinetic energy, $f_{\bar p}(T,\vec{r},t)$, approximately 
satisfies the following transport equation:
\begin{equation}
0=\frac{\partial f_{\bar p}}{\partial t}=
\nabla \cdot (K(T,\vec{r})\nabla f_{\bar p})
-\nabla \cdot (\vec{V_c}(\vec{r})  f_{\bar p})
-2 h \delta(z) \Gamma_{\rm ann} f_{\bar p}+Q(T,\vec{r})\;.
\label{transport-antip}
\end{equation}
The boundary conditions require the solution 
$f_{\bar p}(T,\vec{r},t)$ to vanish at the boundary
of the diffusion zone, which is approximated by a cylinder with 
half-height $L = 1-15~\rm{kpc}$ and radius $ R = 20 ~\rm{kpc}$.

The first term on the right-hand side of the transport equation
is the diffusion
term, which accounts for the propagation through the
tangled Galactic magnetic field.
The diffusion coefficient $K(T,\vec{r})$ is assumed to be constant
throughout the diffusion zone and is parametrized by:
\begin{equation}
K(T)=K_0 \;\beta\; {\cal R}^\delta\;,
\end{equation}
where
$\beta=v/c$ and ${\cal R}$ is the rigidity of the particle, which
is defined as the momentum in GeV per unit charge, 
${\cal R}\equiv p({\rm GeV})/Z$.
The normalization $K_0$ and the spectral index $\delta$
of the diffusion coefficient are related to the properties 
of the interstellar medium and can be determined from the 
flux measurements of other cosmic ray species, mainly from 
the Boron to Carbon (B/C) ratio~\cite{Maurin:2001sj}. 
The second term is the convection term, which accounts for
the drift of charged particles away from the 
disk induced by the Milky Way's Galactic wind. 
It has axial direction and is also assumed to be constant
inside the diffusion region: 
$\vec{V}_c(\vec{r})=V_c\; {\rm sign}(z)\; \vec{k}$.
The third term accounts for antimatter annihilation with rate
$\Gamma_{\rm ann}$, when it interacts with ordinary matter
in the Galactic disk,
which is assumed to be an infinitely thin disk with half-width
$h=100$ pc. The annihilation rate, $\Gamma_{\rm ann}$, is given by
\begin{equation}
\Gamma_{\rm ann}=(n_{\rm H}+4^{2/3} n_{\rm He})
\sigma^{\rm ann}_{\bar p p} v_{\bar p}\;.
\end{equation}
In this expression it has been assumed that the annihilation cross-section
between an antiproton and a helium nucleus is
related to the annihilation cross-section between an
antiproton and a proton by the simple geometrical factor $4^{2/3}$.
On the other hand, $n_{\rm H}\sim 1\;{\rm cm}^{-3}$ is the number
density of Hydrogen nuclei in the Milky Way disk,
$n_{\rm He}\sim 0.07 ~n_{\rm H}$ the number density
of Helium nuclei and $\sigma^{\rm ann}_{\bar p p}$ is
the annihilation cross-section, which is parametrized 
by~\cite{Tan:1983de}:
\begin{eqnarray}
\sigma^{\rm ann}_{\bar p p}(T) = \left\{
\begin{array}{ll}
661\;(1+0.0115\;T^{-0.774}-0.948\;T^{0.0151})\; {\rm mbarn}\;,
 & T < 15.5\;{\rm GeV}~, \\
36 \;T^{-0.5}\; {\rm mbarn}\;, & T \geq 15.5\;{\rm GeV}\,. \\
\end{array} \right. 
\end{eqnarray}
Lastly, $Q(T,\vec{r})$ is the source term of antiprotons from
dark matter annihilations, which was defined in Eq.~(\ref{eq:source}). 
In this equation, energy losses, reacceleration effects and 
non-annihilating interactions of antimatter in the Galactic disk have been 
neglected. The ranges of the astrophysical parameters that are
consistent with the B/C ratio and that produce the maximal, median and minimal
antiproton fluxes are listed in Table \ref{tab:param-antiproton}.

\begin{table}[t]
\begin{center}
\begin{tabular}{|c|cccc|}
\hline
Model & $\delta$ & $K_0\,({\rm kpc}^2/{\rm Myr})$ & $L\,({\rm kpc})$
& $V_c\,({\rm km}/{\rm s})$ \\
\hline 
MIN & 0.85 & 0.0016 & 1 & 13.5 \\
MED & 0.70 & 0.0112 & 4 & 12 \\
MAX & 0.46 & 0.0765 & 15 & 5 \\
\hline
\end{tabular}
\caption{\label{tab:param-antiproton}
Astrophysical parameters compatible with the B/C ratio that 
yield the minimal (MIN), median (MED) and maximal (MAX) flux of antiprotons.}
\end{center}
\end{table}

Finally, the interstellar flux of primary antiprotons at the Solar System
from dark matter annihilation is given by:
\begin{equation}
\Phi^{\rm{IS}}_{\bar p}(T) = \frac{v}{4 \pi} f_{\bar p}(T,r_\odot),
\label{flux}
\end{equation}
where $v$ is the antiproton velocity\footnote{We have cross-checked our results using the  interpolating functions presented in \cite{Cirelli:2010xx}.}.

However, this is not the antiproton flux measured by experiments, 
which is affected by solar modulation.
In the force field approximation~\cite{solar-modulation} 
the effect of solar modulation can be included
by applying the following simple formula that relates 
the antiproton flux at the top of the Earth's atmosphere and
the interstellar antiproton flux~\cite{perko}:
\begin{equation}
\Phi_{\bar p}^{\rm TOA}(T_{\rm TOA})=
\left(
\frac{2 m_p T_{\rm TOA}+T_{\rm TOA}^2}{2 m_p T_{\rm IS}+T_{\rm IS}^2}
\right)
\Phi_{\bar p}^{\rm IS}(T_{\rm IS}),
\end{equation}
where $T_{\rm IS}=T_{\rm TOA}+\phi_F$, with
$T_{\rm IS}$ and $T_{\rm TOA}$ being the antiproton kinetic energies 
at the heliospheric boundary and at the top of the Earth's atmosphere,
respectively, and $\phi_F$ being the solar modulation parameter,
which varies between 500 MV and 1.3 GV over the eleven-year solar
cycle. Since experiments are usually undertaken near
solar minimum activity, we will choose $\phi_F=500$ MV 
for our numerical analysis.

\begin{figure}
\hspace*{-0.6cm}
\begin{tabular}{ll}
 \includegraphics[width=0.525\textwidth]{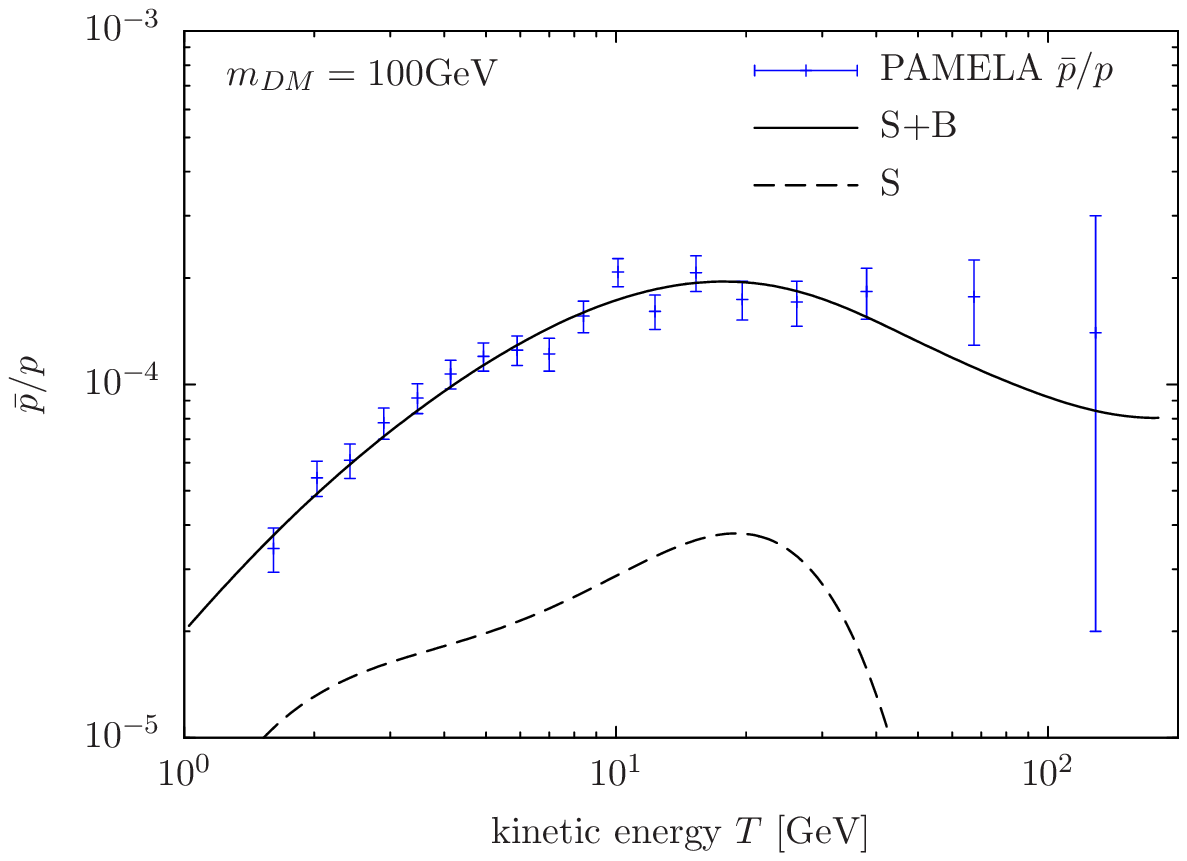}
 &\hspace*{-0.75cm} \includegraphics[width=0.525\textwidth]{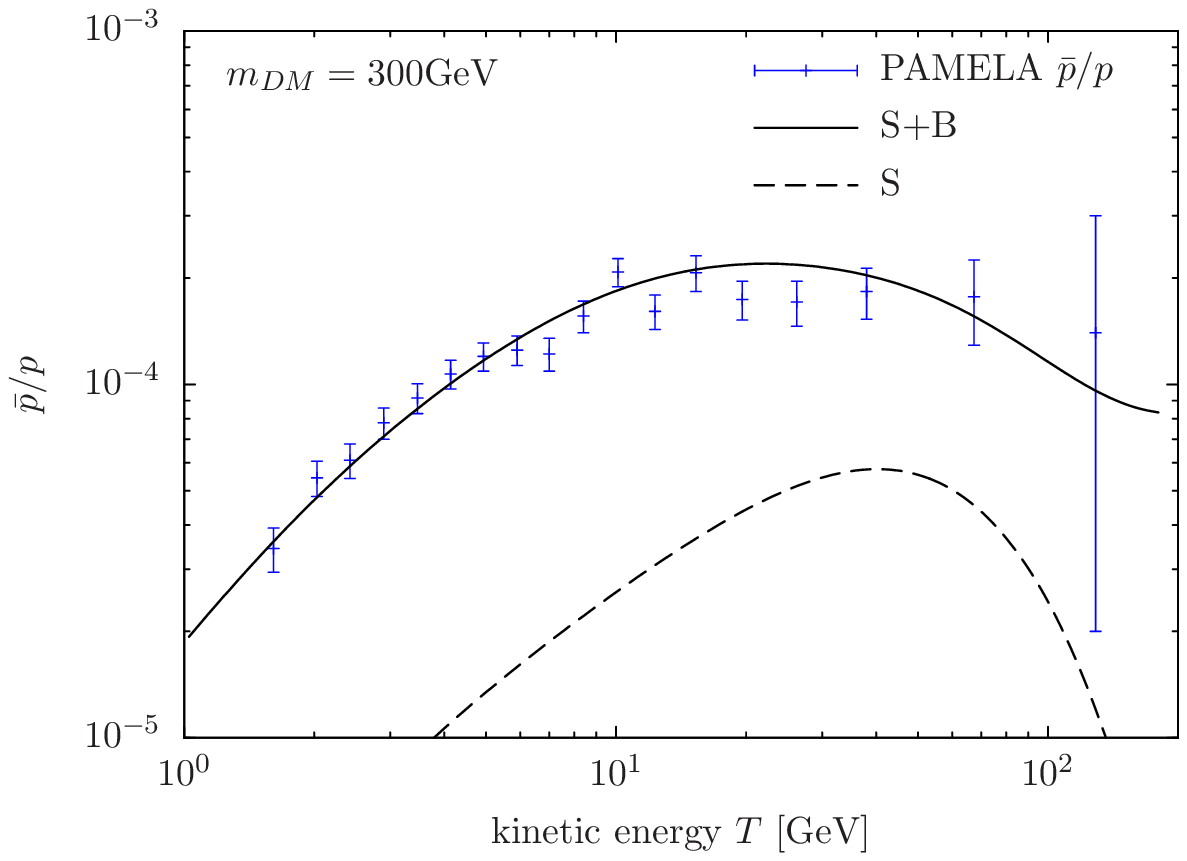} \\
 \includegraphics[width=0.525\textwidth]{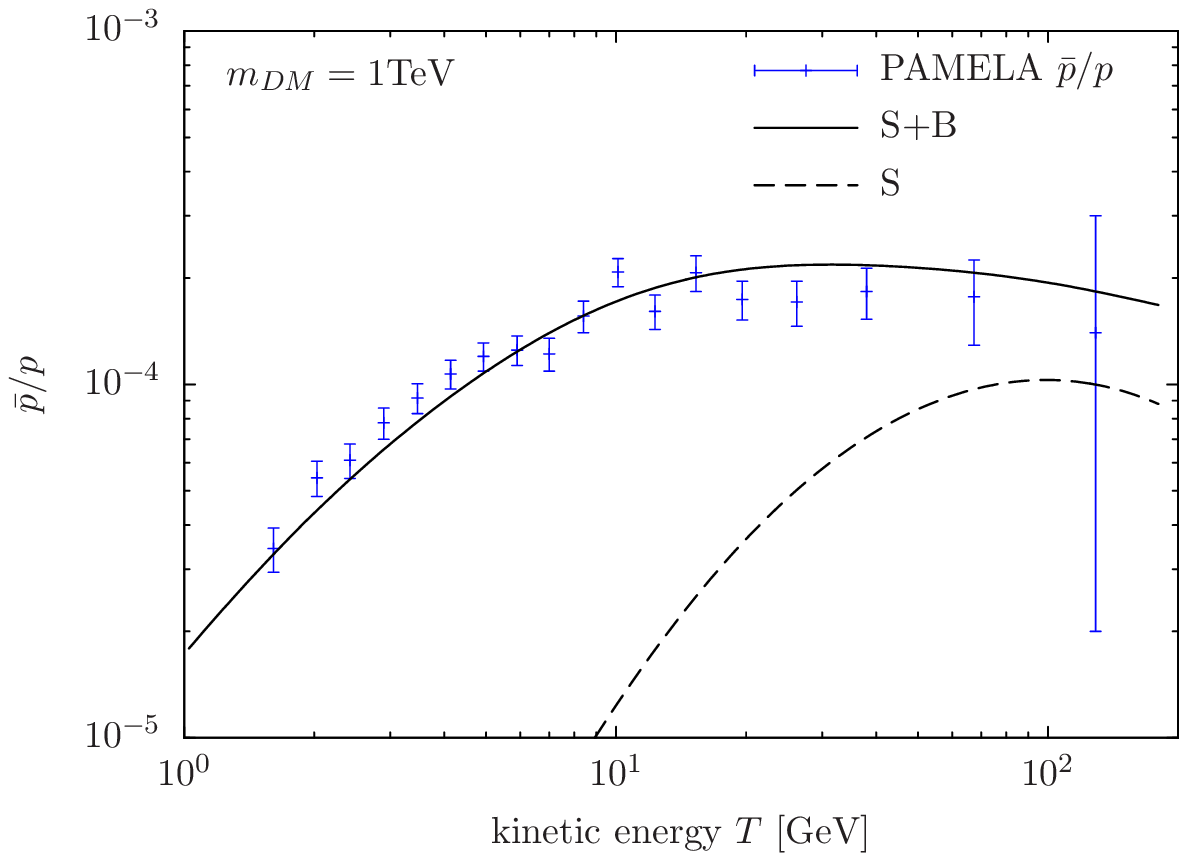}
 &\hspace*{-0.75cm} \includegraphics[width=0.525\textwidth]{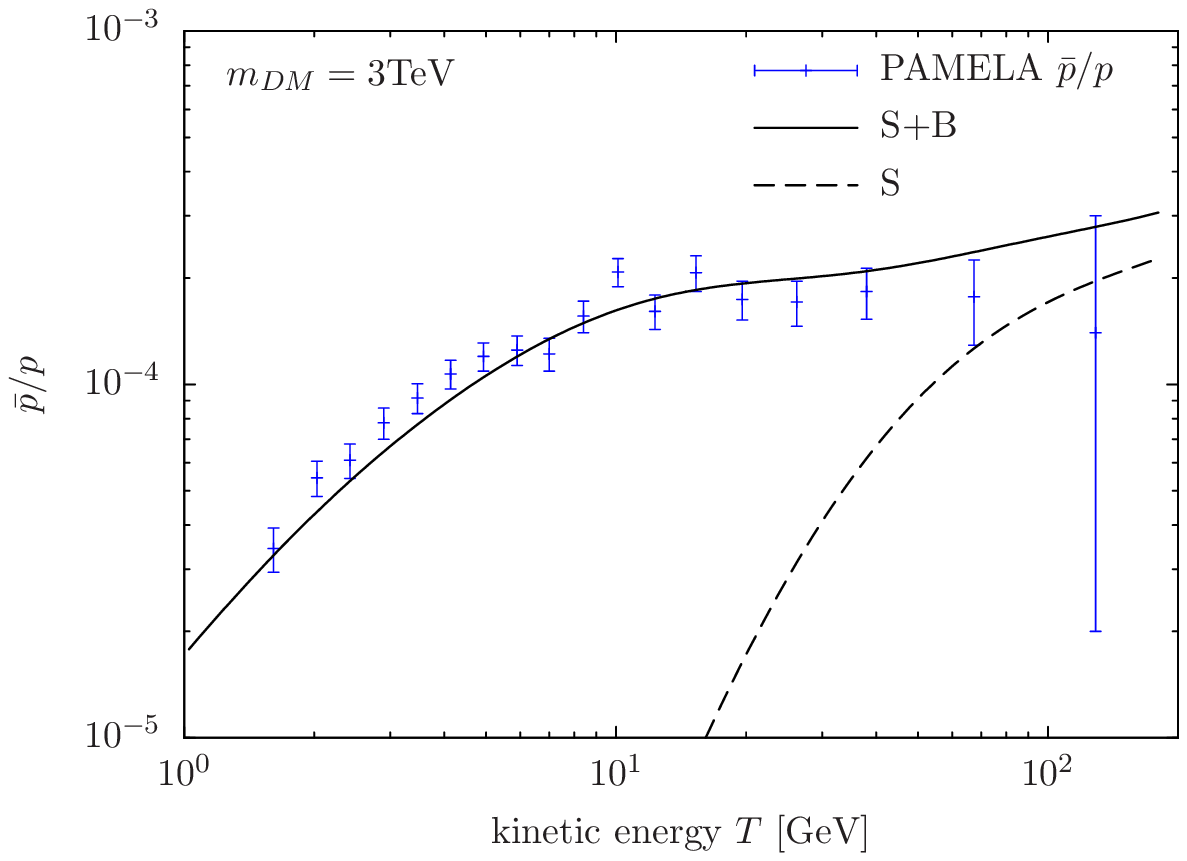}
\end{tabular}
 \caption{\label{fig:pbarp} Antiproton-to-proton ratio obtained from electroweak IB, for dark matter masses 100GeV (upper left), 300GeV (upper right), 1TeV (lower left) and 3TeV (lower right). Also shown are the PAMELA data for the $\bar p/p$ ratio \cite{Adriani:2010rc}. The two lines correspond to the signal allowed at 95\%C.L. (dashed), and to signal+background (solid), respectively. The antiproton background is taken from \cite{Donato:2001ms}.}
\end{figure}

In order to obtain constraints on the dark matter annihilation cross-section, we compute
the antiproton-to-proton ratio
\begin{equation}
 \frac{\bar p}{p} \equiv \frac{ \Phi_{\bar p}^{\rm sig} +  \Phi_{\bar p}^{\rm bkg} }{\Phi_p} \,,
\end{equation}
using the proton flux of \cite{Bringmann:2006im}. Furthermore, in order to obtain a conservative exclusion bound we adopt the minimal value for the antiproton background as discussed in \cite{Donato:2001ms}.

In Fig.\,\ref{fig:pbarp} we show the resulting $\bar p/p$ ratio for various dark matter masses, together with the PAMELA data \cite{Adriani:2010rc}. The data can be described consistently by the background only. Therefore, the measurements can be used to set an upper limit on the contribution $\Phi_{\bar p}^{\rm sig}$ to the antiproton flux originating from dark matter annihilations, which in turn yields an upper bound on the corresponding annihilation cross-section.

\begin{figure}
\includegraphics{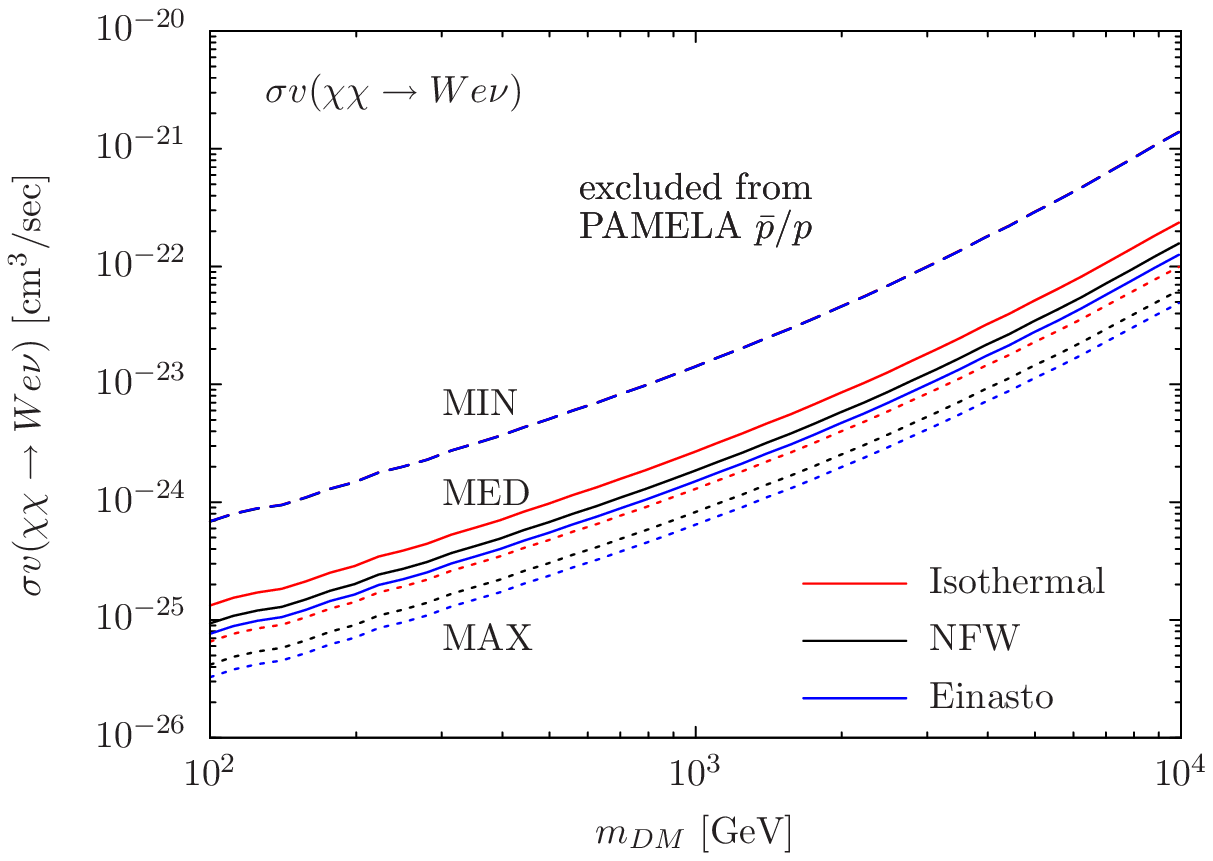}
\includegraphics{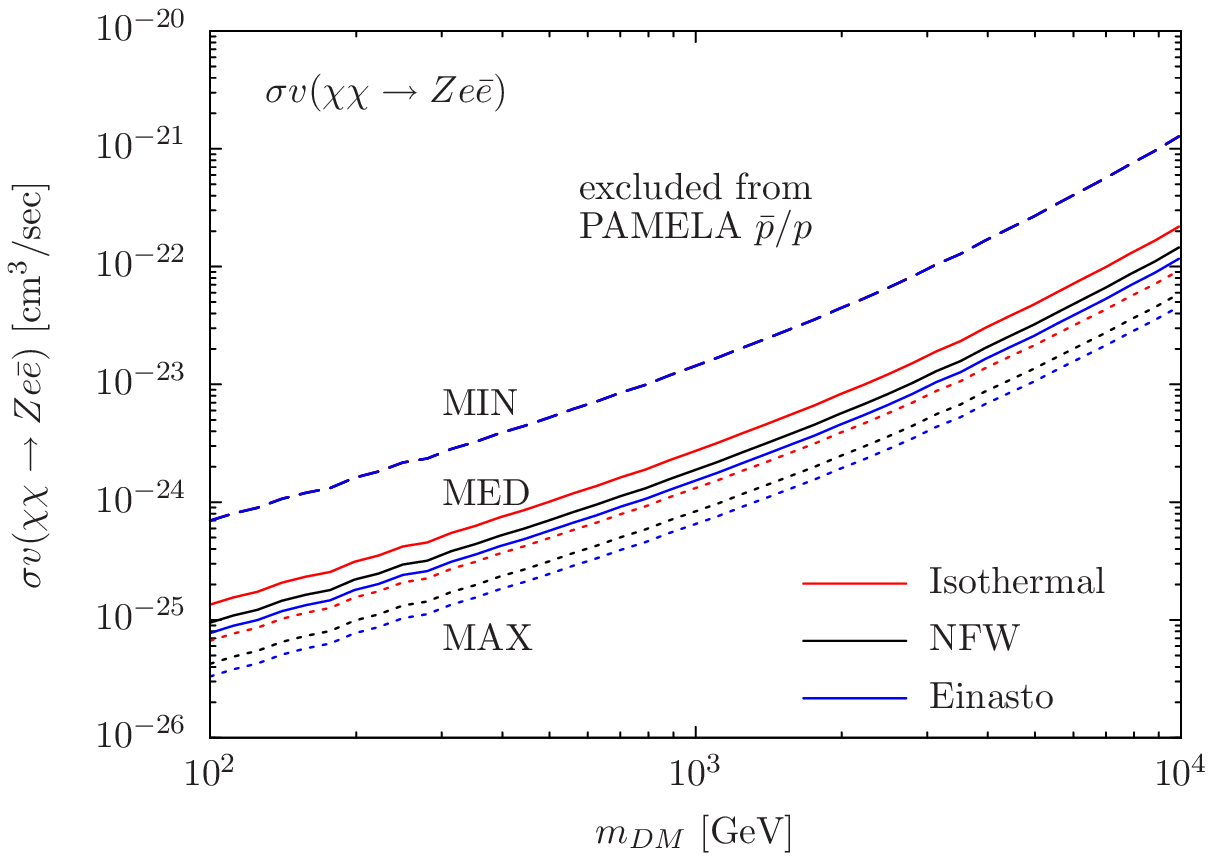}
 \caption{\label{fig:exclusionPlot1} Upper bounds on the electroweak IB cross-sections $\sigma v(2\to 3)$ obtained from the PAMELA data \cite{Adriani:2010rc} of the cosmic antiproton-to-proton ratio. The plots correspond to the constraints on the individual electroweak IB processes $\chi\chi\to W e\nu$ (where $W e\nu\equiv W^- \bar e\nu + W^+ e\bar \nu$, upper), and $\chi\chi\to Z e\bar e$ (lower) at 95\%C.L. The bounds on $\chi\chi\to Z \nu\bar \nu$ are identical because the spectrum of $Z$ bosons per annihilation is the same as for $Ze\bar e$. We have used $m_{\eta^\pm}/m_{DM}=1.5$ and  $m_{\eta^0}^2-m_{\eta^\pm}^2= v_{EW}^2$. Dashed lines correspond to the MIN, solid to MED and dotted to MAX propagation models. The bounds obtained assuming an isothermal dark matter profile are shown in red, NFW in black, and Einasto in blue.}
\end{figure}

We show in Fig.\,\ref{fig:exclusionPlot1} our results for the upper limits on the cross-sections of the individual electroweak IB processes obtained from the PAMELA $\bar p/p$ data \cite{Adriani:2010rc} using a $\chi^2$-test at $95\%$C.L. In order to take the astrophysical uncertainties into account, we compute limits for the three propagation models and three halo profiles discussed above. However, we emphasize that the limits on the individual channels are rather robust from the particle physics side. The reason is that they just depend on the shape of the antiproton spectrum, which turns out not to depend strongly on e.g. the scalar masses, as long as $m_{\eta^i}/m_{DM}\sim\mathcal{O}(1)$. For the MED propagation model the limits lie in the range $\sigma v\lesssim 10^{-25}\cm^3/\sec$ for $m_{DM}=100\GeV$ and $10^{-24}\cm^3/\sec$ for $m_{DM}=1\TeV$ for all the electroweak IB processes.

\begin{figure}
\hspace*{-0.5cm}
\begin{tabular}{ll}
 \includegraphics[width=0.525\textwidth]{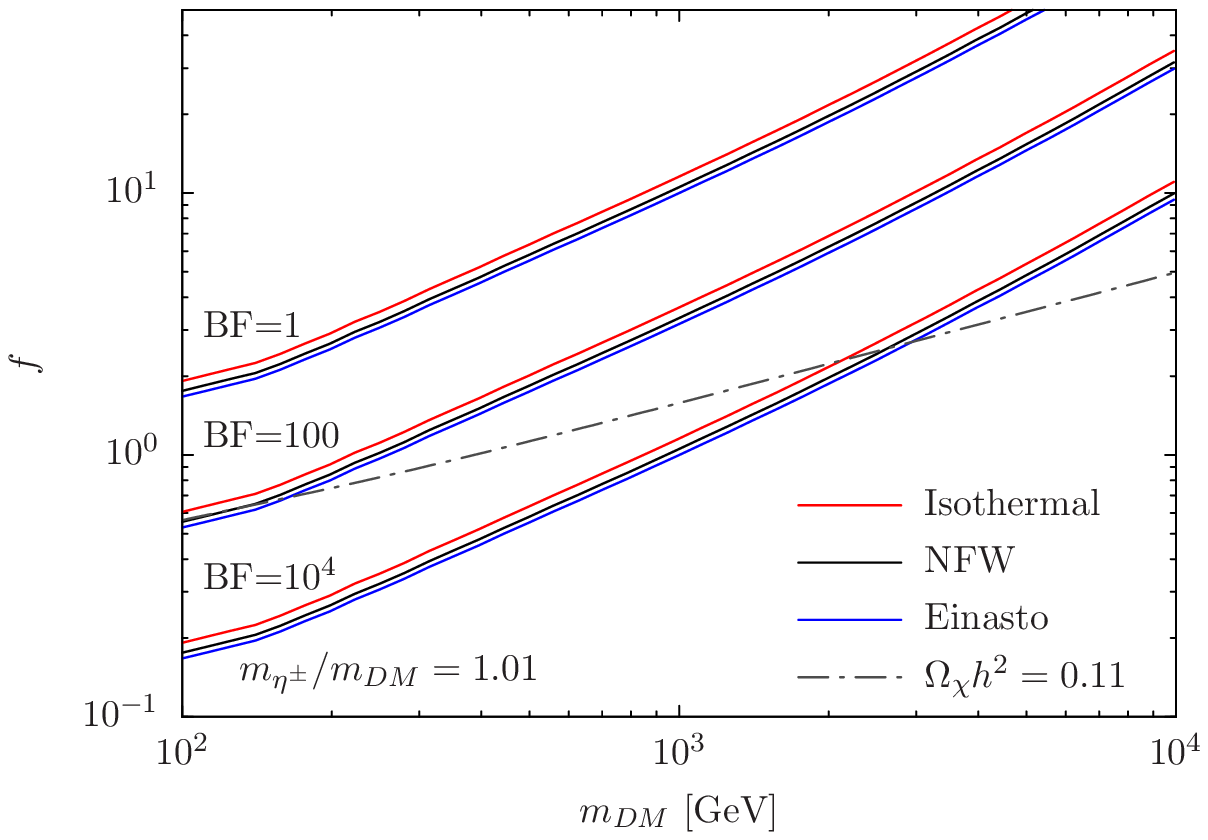}
 &\hspace*{-0.65cm} \includegraphics[width=0.525\textwidth]{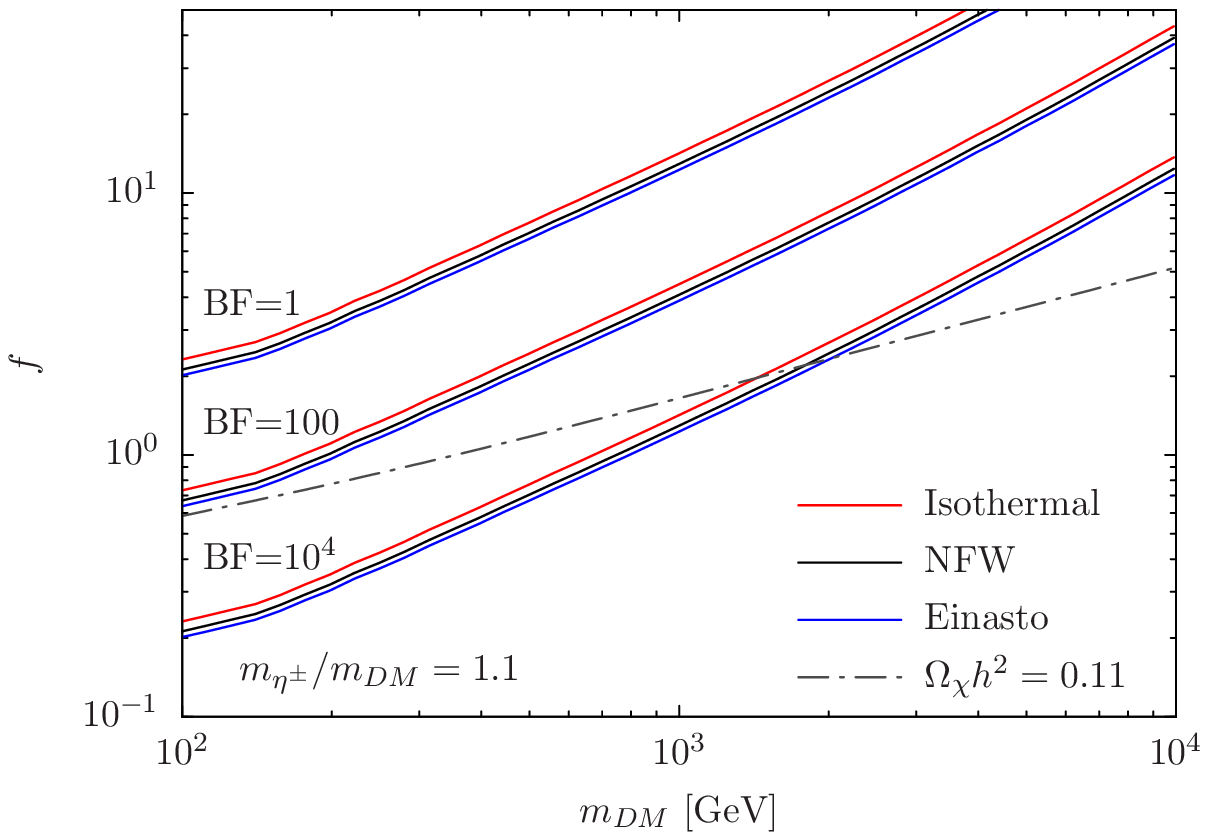} \\
 \includegraphics[width=0.525\textwidth]{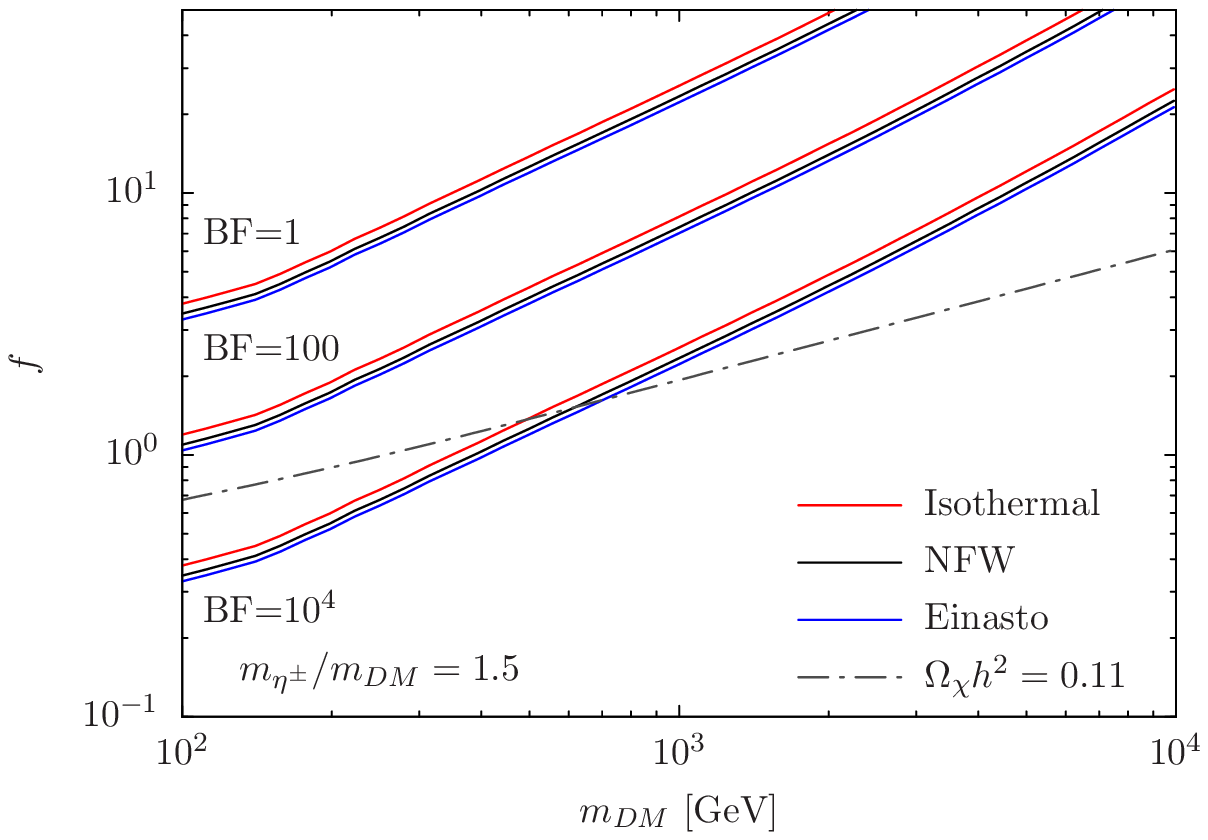}
 &\hspace*{-0.65cm} \includegraphics[width=0.525\textwidth]{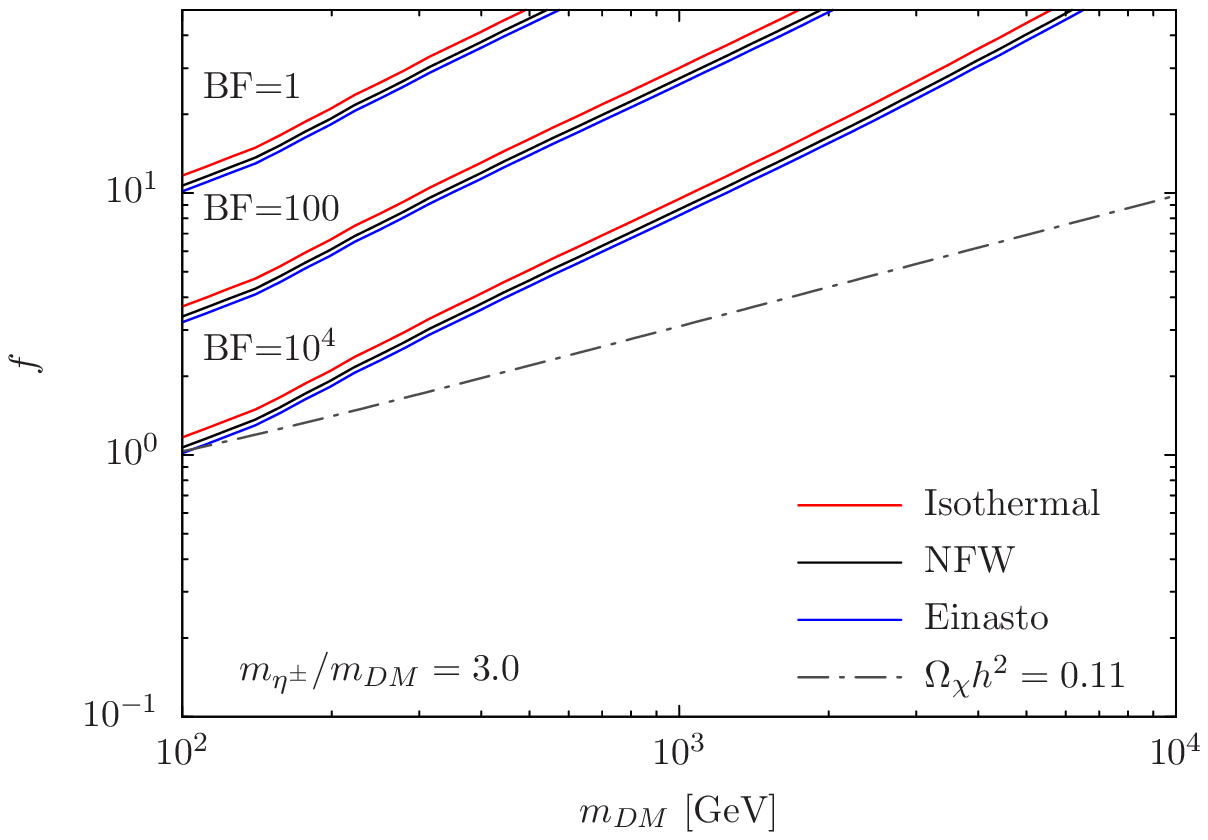}
\end{tabular}
 \caption{\label{fig:exclusionPlot2} Upper bounds on the dark matter coupling $f$ inferred from the antiproton constraints on the electroweak IB processes (95\%C.L.) for $m_{\eta^0}^2-m_{\eta^\pm}^2=v_{EW}^2$ and different choices for the ratio $m_{\eta^\pm}/m_{DM}=1.01,1.1,1.5,3.0$. Also shown is the coupling for which thermal freeze-out would yield a relic density compatible with WMAP within the toy model \cite{Cao:2009yy} (dot-dashed), as well as the effect of a hypothetical boost factor $BF$ in the Milky Way on the bound. The bounds correspond to the MED propagation model.}
\end{figure}

In order to compare our results with other constraints, like e.g. the relic density, it is convenient to convert the exclusion limits on the cross-sections into constraints on the parameter space of the toy model.
All the cross-sections relevant for dark matter annihilation are proportional to $f^4$, where $f$ characterizes the coupling strength of the dark matter particle to the Standard Model. Thus, the exclusion limits can be converted into upper limits on $f$, by using the total cross-sections derived in Section\,\ref{sec:pp}.

The resulting upper limits on the coupling $f$ depend strongly on the ratio $m_{\eta^\pm}/m_{DM}$, because of the enhancement of the total IB cross-sections when $m_{\eta^\pm}/m_{DM}\to 1$. The results for several values of the ratio are shown in Fig.\,\ref{fig:exclusionPlot2}. For comparison, we also show the coupling for which thermal freeze-out would yield the correct relic density within the toy model \cite{Cao:2009yy}, as well as the exclusion bounds scaled with hypothetical boost factors $BF$ in the Milky Way. For $BF=1$, the upper bounds on the coupling from antiproton constraints are compatible with the couplings that yield the correct relic density, for all dark matter masses and all values of $m_{\eta^\pm}/m_{DM}$ considered in Fig.\,\ref{fig:exclusionPlot2}. For sub-TeV dark matter masses, the antiproton constraints become relevant for very large values $BF\sim 10^2$ of the boost factor, when assuming that $\chi$  was produced thermally and constitutes the dominant component of dark matter. In that case, the antiproton constraints arising from electroweak IB may be considered as conservative upper bounds on the size of the coupling $f$.

\begin{figure}
\hspace*{-0.5cm}
\begin{tabular}{ll}
 \includegraphics[width=0.525\textwidth]{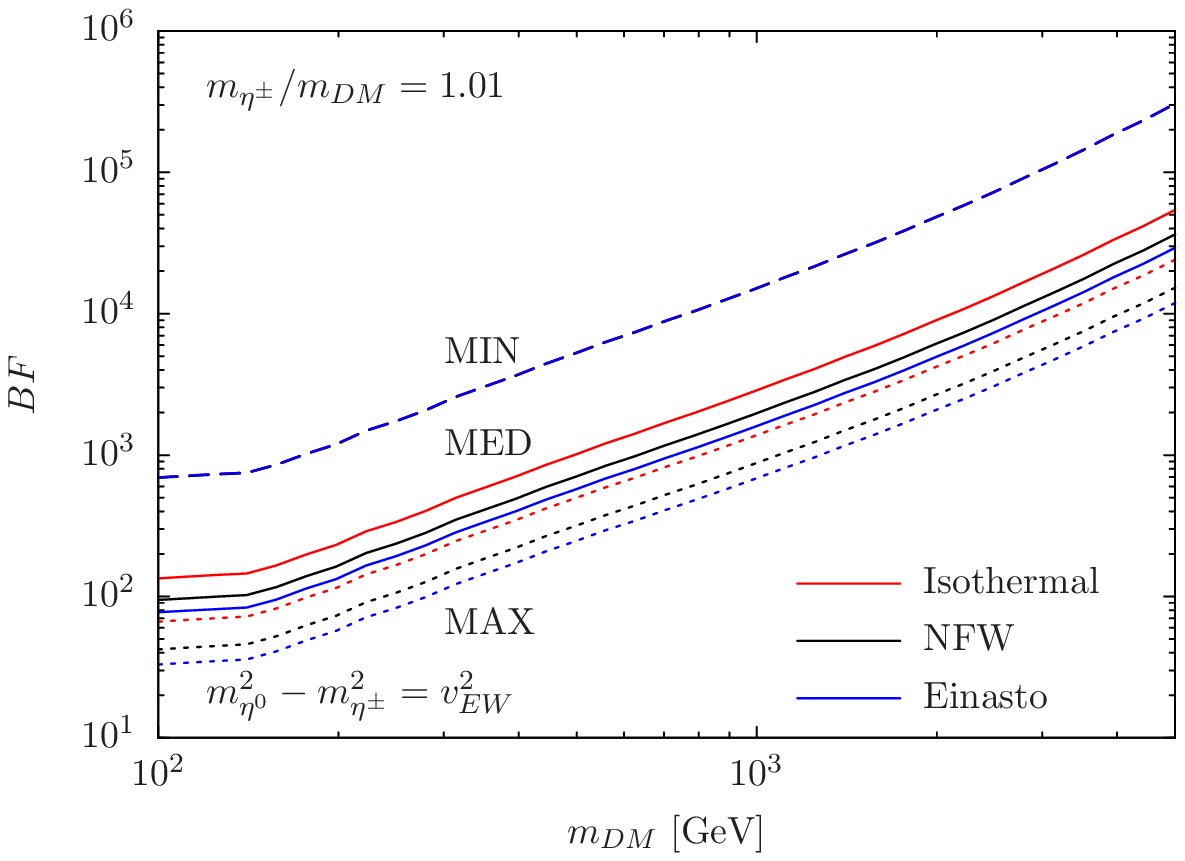}
 &\hspace*{-0.65cm} \includegraphics[width=0.525\textwidth]{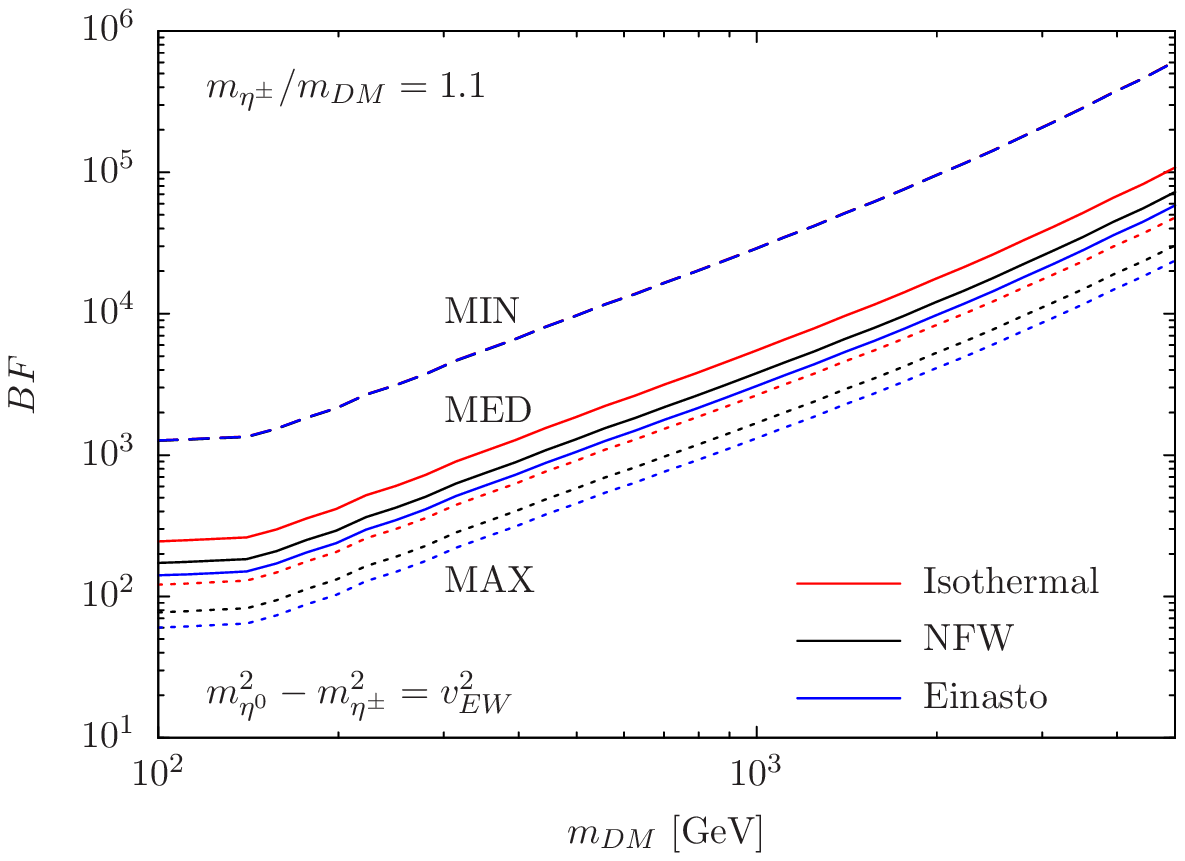} \\
 \includegraphics[width=0.525\textwidth]{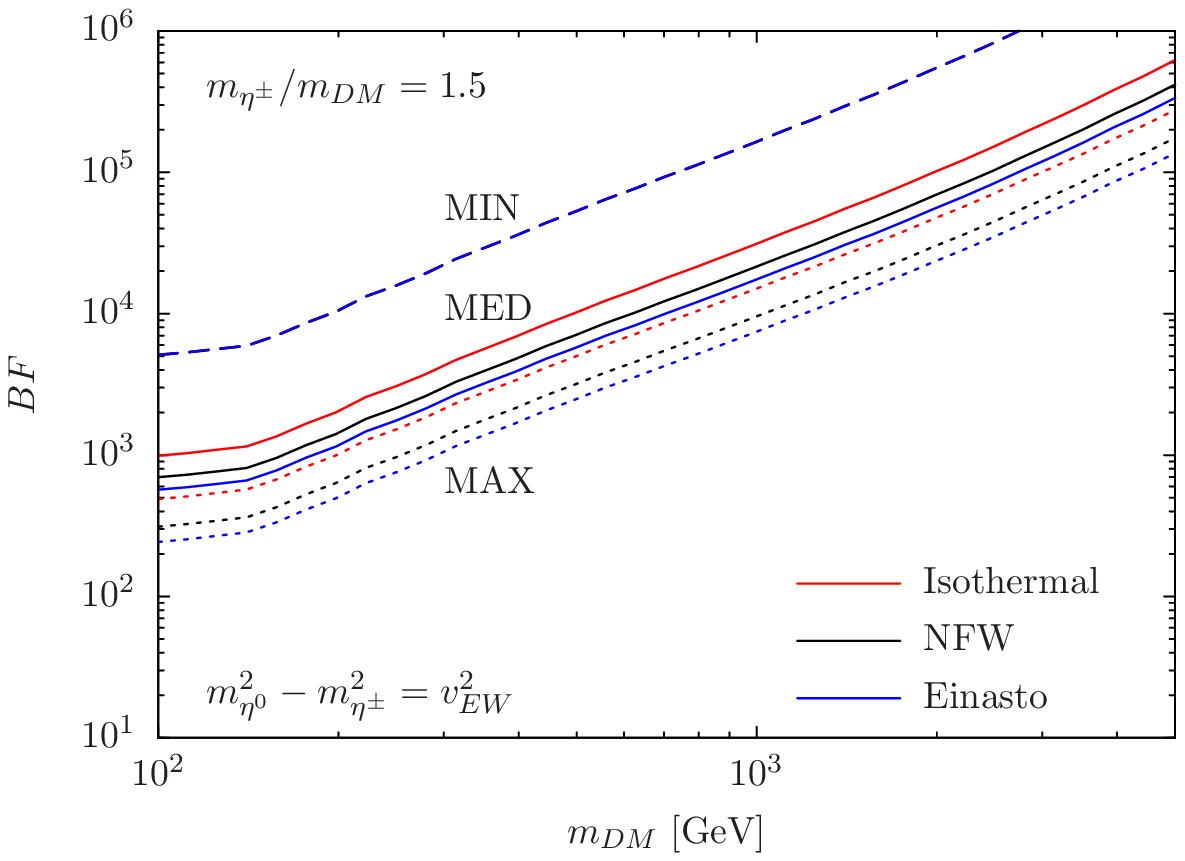}
 &\hspace*{-0.65cm} \includegraphics[width=0.525\textwidth]{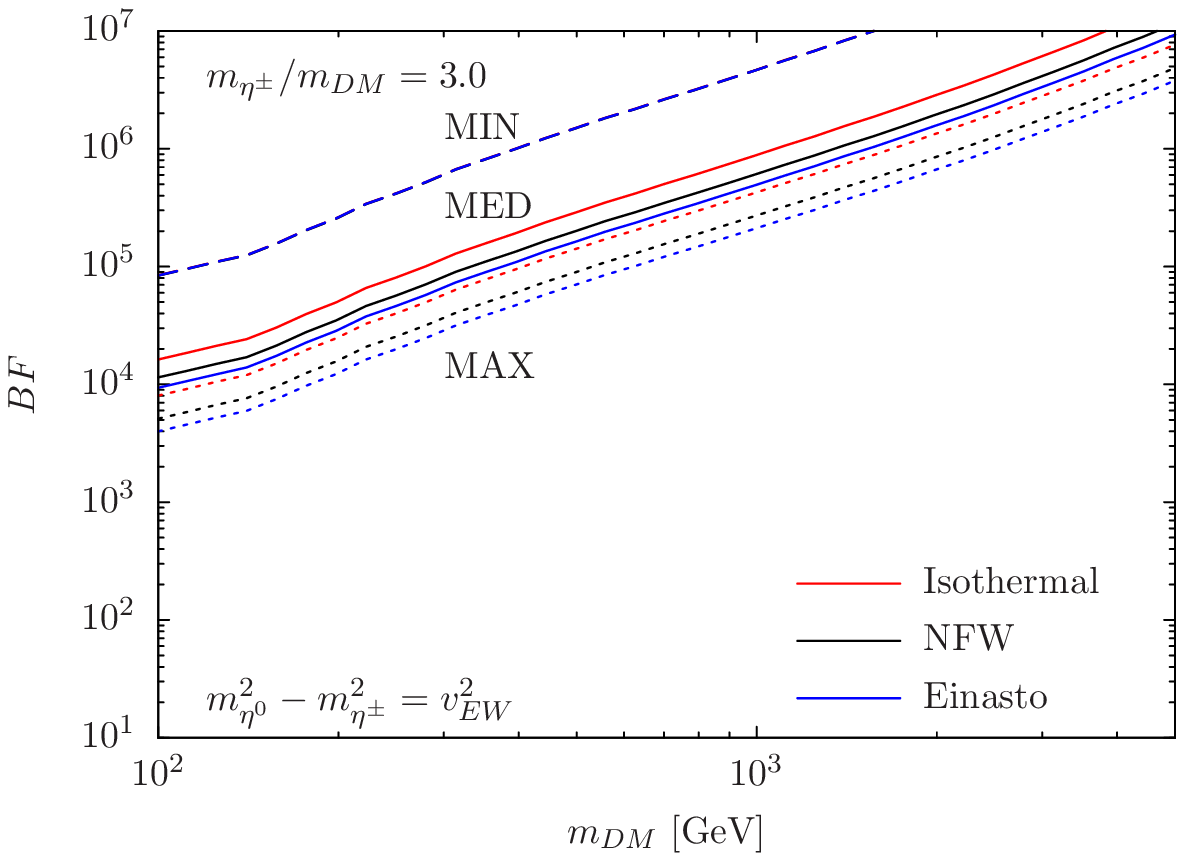}
\end{tabular}
 \caption{\label{fig:exclusionPlot3} Upper bounds on the boost factor $BF$ in the Milky Way inferred from the antiproton constraints on the electroweak IB processes (95\%C.L.) for $m_{\eta^0}^2-m_{\eta^\pm}^2=v_{EW}^2$ and different choices for the ratio $m_{\eta^\pm}/m_{DM}=1.01,1.1,1.5,3.0$. The coupling $f$ is fixed for each dark matter mass from requiring a relic density from thermal production $\Omega h^2=0.11$, see Eq.\,(\ref{eq:omegaDM}).}
\end{figure}

\begin{figure}
\hspace*{-0.6cm}
\begin{tabular}{ll}
 \includegraphics[width=0.525\textwidth]{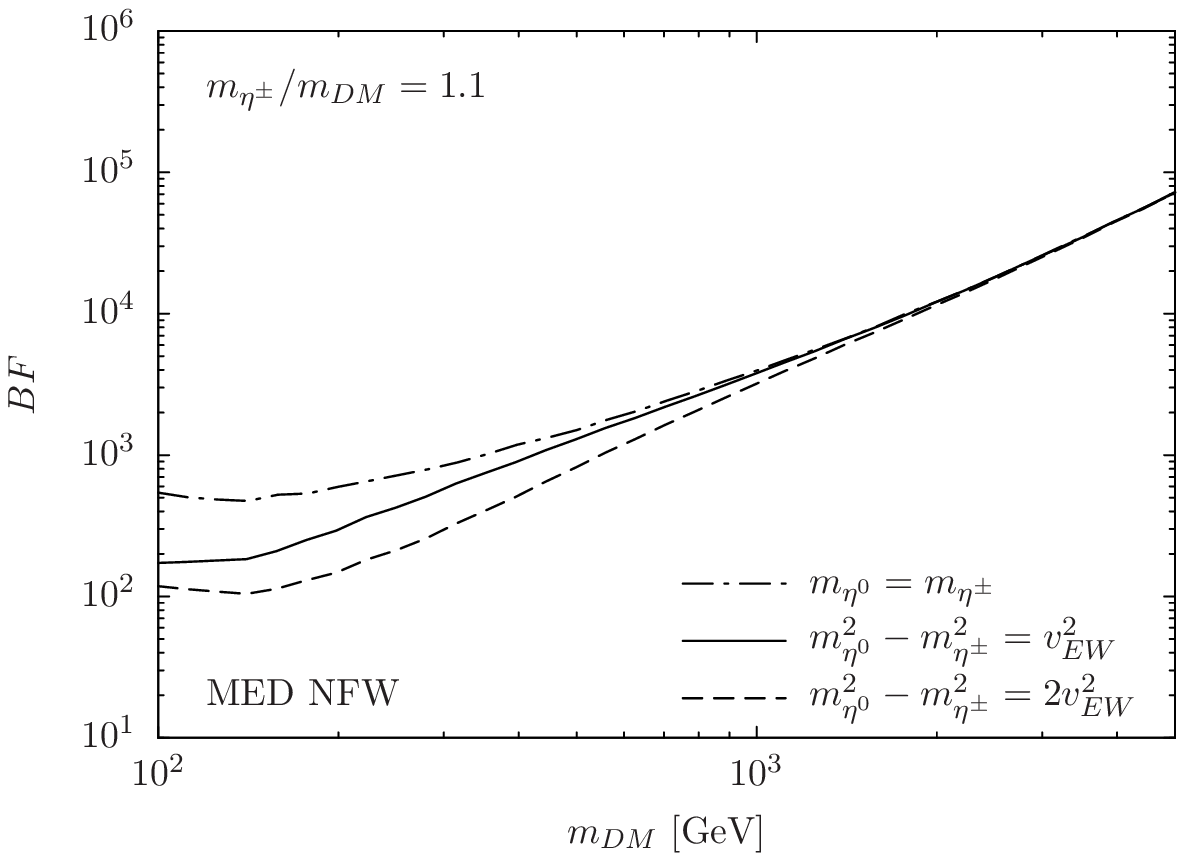}
 &\hspace*{-0.75cm} \includegraphics[width=0.525\textwidth]{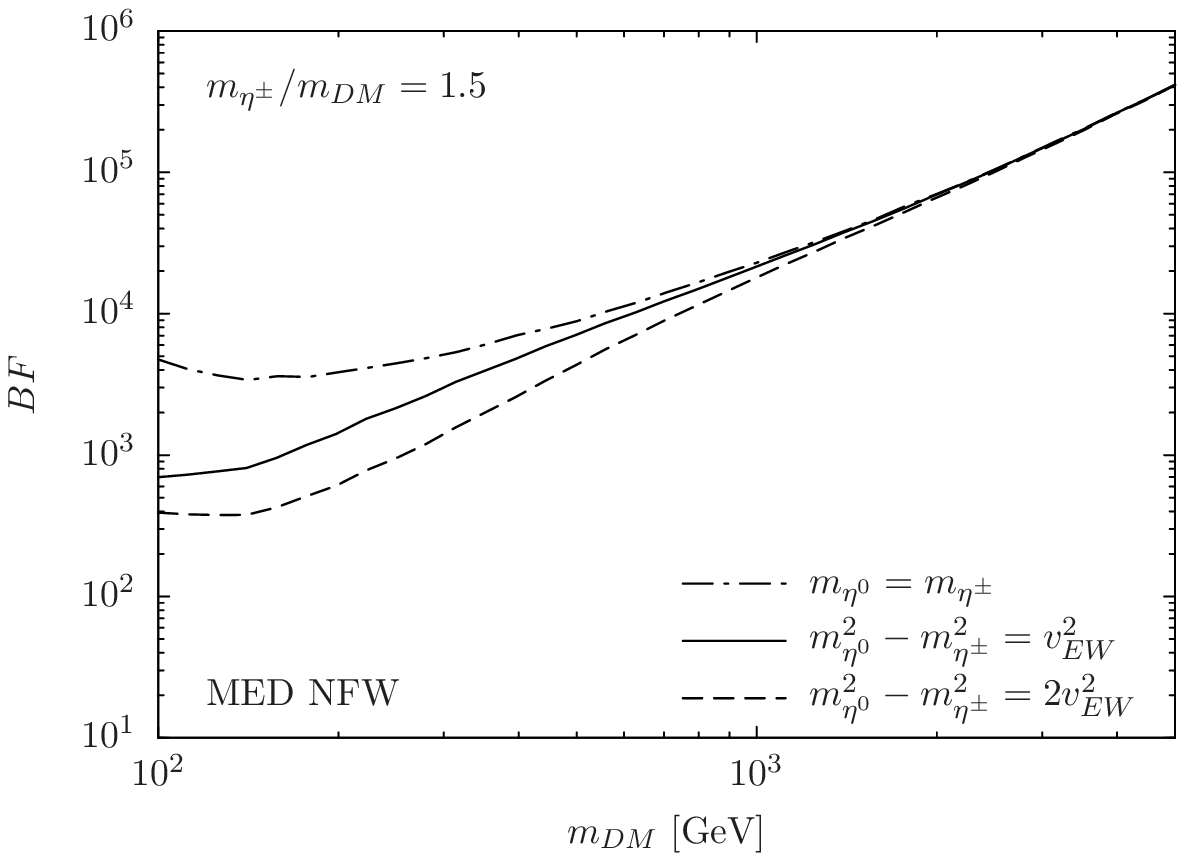}
\end{tabular}
 \caption{\label{fig:exclusionPlot4} Upper bounds on the boost factor $BF$ in the Milky Way inferred from the antiproton constraints on electroweak IB processes (95\%C.L.) for $m_{\eta^\pm}/m_{DM}=1.1$ (left) and $1.5$ (right) and different choices for the mass-splitting $m_{\eta^0}^2-m_{\eta^\pm}^2$. The coupling $f$ is fixed for each dark matter mass from requiring a relic density from thermal production $\Omega h^2=0.11$, see Eq.\,(\ref{eq:omegaDM}). The bounds correspond to the MED propagation model and the NFW halo profile.}
\end{figure}

Nevertheless, the constraints can be used to evaluate the possibility of observing signals from dark matter annihilations in other channels, in particular gamma rays or positrons, with present or planned instruments.
For example, it has been argued that the positron excess observed in the PAMELA data \cite{Adriani:2011xv,Adriani:2008zr}, could be explained by dark matter annihilations into leptons. In particular, an enhancement of the annihilation into leptons due to internal bremsstrahlung has been considered \cite{Bergstrom:2008gr}. Note that in this context boost factors of the order $\sim 10^4$ have been invoked in the literature. Despite a lack of explanation for astrophysical boost factors of this size, we note that the constraints coming from antiproton production due to electroweak IB independently restrict this interpretation of the PAMELA positron excess. Also, since the production mechanism of dark matter is unknown, it is worthwhile to investigate the viability of observably large dark matter annihilation cross-sections. In this sense, the boost factor can serve as a reference value in order to compare the cross-relations between prospects in different annihilations channels.

Detecting a gamma-ray signal from internal bremsstrahlung at MAGIC II and CTA, for instance, typically requires boost factors in the target (e.g. the dwarf galaxies Draco or Willman 1) of the order of $BF\gtrsim 10^3-10^4$ according to Ref.\,\cite{Bringmann:2008kj}. We stress that antiprotons from electroweak IB indeed yield relevant constraints for boost factors in the Milky Way of this size.

If one assumes that the coupling $f$ is fixed such that the correct relic density is produced in the early universe, then it is possible to obtain upper limits on the boost factors $BF$ that are compatible with antiproton constraints. The corresponding exclusion plots are shown in Fig.\,\ref{fig:exclusionPlot3}, again for several values of the ratio $m_{\eta^\pm}/m_{DM}$, and assuming a generic scalar mass-splitting $m_{\eta^0}^2-m_{\eta^\pm}^2=v_{EW}^2$. The results obtained when assuming $m_{\eta^0}=m_{\eta^\pm}$ and $m_{\eta^0}^2-m_{\eta^\pm}^2=2v_{EW}^2$, respectively, are shown in Fig.\,\ref{fig:exclusionPlot4} for comparison. We find that, for the case $m_{\eta^\pm}/m_{DM}\sim 1$, which yields a particularly strong gamma (and possibly positron) signal from electromagnetic IB, also the constraints from electroweak IB become quite stringent in comparison, of the order $BF\lesssim 10^2-10^3$.

\section{Implications for realistic models: the case of the lightest neutralino}\label{sec:neutralino}

\begin{table}[t]
\begin{center}
\begin{tabular}{|c|c|ccc|}
\hline
Model &  $m_{DM}$ &   & $BF$ ($\bar p/p$) &    \\
      &  [GeV]    & MIN & MED & MAX  \\
      &           & $2\to2/3\,(2\to 3)$ & $2\to2/3\,(2\to 3)$ & $2\to2/3\,(2\to 3)$  \\
\hline
BM2    & 453 & $< 4.4\cdot10^4$ ($9.6\cdot10^5$) & $< 5900$ ($1.3\cdot10^5$) & $< 2700$ ($5.8\cdot10^4$)  \\
BM3    & 234 & $< 1.0\cdot10^4$ ($9.9\cdot10^4$) & $< 1500$ ($1.3\cdot10^4$) & $< 660$  ($6100$)          \\
BM$J'$ & 316 & $< 2400$         ($2.6\cdot10^5$) & $< 330$  ($3.5\cdot10^4$) & $< 150$  ($1.6\cdot10^4$)  \\
BM$I'$ & 141 & $< 82$           ($5.1\cdot10^4$) & $< 11$   ($6900$)         & $< 5.0$  ($3100$)          \\
\hline
\end{tabular}
\caption{\label{tab:bf-mssm} Upper limits on the boost factor $BF$ in the Milky Way obtained
from the PAMELA $\bar p/p$ data \cite{Adriani:2010rc} for several MSSM benchmark points (see \cite{Bringmann:2007nk,Bringmann:2008kj} for details) at $95\%$C.L.  We have taken into account the antiproton production from $2\to 2$ processes with hadronic final states as well as $2\to 3$ processes that are enhanced due to electroweak IB. The values in parentheses are the upper limits on the boost factor that would be obtained when taking only the $2\to 3$ processes into account. The bounds correspond to the MIN, MED and MAX propagation models and the NFW halo profile.
}
\end{center}
\end{table}

In the previous section we have derived bounds on the $2\rightarrow 3$ annihilation cross-section and the corresponding boost factor for a toy model of
leptophilic dark matter which only couples to the electron doublet and an extra scalar doublet. However, in more realistic models the WIMP dark matter
particle also couples to the quarks and the gauge bosons. As a consequence, the antiproton production in $2\rightarrow 2$ processes can be significant or
even dominate over the production in the fragmentation of the weak gauge bosons from the $2\rightarrow 3$ processes which we have analyzed in this paper. It
is then apparent that the bounds derived in this paper for the boost factors can be used as conservative bounds for any realistic model.

In order to estimate how are these conservative bounds modified when considering realistic models, we have studied the very popular scenario where the dark matter particle is the lightest neutralino. We have analyzed the antiproton production in the supersymmetric benchmark points BM2, BM3, BMJ', BMI' (using~\cite{Djouadi:2002ze,Belanger:2004yn}); the points BM2 and BM3 lie in the coannihilation region, BMJ' in coannihilation tail and BMI' in the bulk region. All of them are characterized by a large cross-section for the internal electromagnetic bremsstrahlung processes. Furthermore, BM3, BMJ' and BMI' were used in \cite{Bringmann:2008kj} to evaluate the prospects to observe a gamma-ray signal in dark matter annihilations at MAGIC II or at the projected CTA. Our results are shown in Table \ref{tab:bf-mssm}, where we list the upper limit on the boost factor in the Milky Way from requiring no excess in the PAMELA $\bar p/p$ data at the 95\% C.L., assuming the NFW halo profile and for the MIN, MED and MAX propagation models.We also show in parentheses the upper bound corresponding to just considering the antiproton production in the electroweak bremsstrahlung. From the table it follows that in realistic models the upper bounds on the boost factor are at least one order of magnitude stronger than the conservative bounds derived in this paper, and in some cases even three orders of magnitude stronger.

It is interesting to compare our upper bounds on the boost factor in the Milky way with the lower bounds on the boost factor derived in
\cite{Bringmann:2008kj} for the Draco dwarf spheroidal galaxy, from requiring the detection of a gamma-ray signal with a $5\sigma$ significance at MAGIC II
or at the projected CTA with 30 hours of observation. The observation of such a signal at MAGIC II requires a boost factor in Draco larger than $\sim 10^4$,
and at the CTA larger than  $\sim 10^3$. In contrast, we obtain that the non-observation of an excess in the PAMELA $\bar p/p$ data implies, for the same
benchmark point, a boost factor in the Milky Way smaller than $\sim 10^2-10^3$ for the MED propagation model. Therefore, the observation of a gamma-ray
signal from Draco at MAGIC II will imply the existence of a boost factor in Draco much different to the boost factor in the Milky way.
Furthermore, the supersymmetric benchmark point BM3 has been discussed in \cite{Bergstrom:2008gr} as a possible explanation of the PAMELA positron excess,
provided the boost factor in the Milky way is $\sim 3\times 10^4$. This interpretation is then in tension with the upper bounds for BM3 shown in Table
\ref{tab:bf-mssm}, unless there is a mechanism which boosts the positron signal significantly more than the antiproton signal.

\section{Conclusions}

We have analyzed the antiproton production in the annihilation of dark matter Majorana particles into two leptons and one weak gauge boson. We are motivated by the possibility of observing a gamma-ray signal from the internal electromagnetic bremsstrahlung process $\chi\chi\rightarrow f \bar f \gamma$, which produces a characteristic feature in the gamma-ray spectrum and which may constitute a smoking-gun for dark matter detection. We have argued that if such a signal is observed in gamma-rays, the $SU(2)_L\times U(1)_Y$ gauge invariance of the Standard Model automatically implies annihilations into two fermions and one weak gauge boson. Then, the fragmentation of the fermions and weak gauge bosons produces an antiproton flux, which is severely constrained by the PAMELA measurements of the cosmic antiproton-to-proton flux ratio. As a consequence, the prospects to observe a gamma-ray signal from internal electromagnetic bremsstrahlung at  currently operating IACTs and at the projected CTA is constrained by the PAMELA measurements on the antiproton-to-proton ratio. 

Concretely, we have studied a model where the dark matter particle is a Majorana fermion which couples only to the electron doublet and to an extra scalar doublet, heavier than the dark matter particle. We have calculated analytically and numerically the cross-sections for the $2\rightarrow 2$ and $2\rightarrow 3$ processes in all relevant scenarios, which we catalogue according to the level of degeneracy between the masses of the charged and the neutral components of the scalar doublet, and the level of degeneracy between the masses of the charged scalar and the dark matter particle. For all these cases, we have also calculated the expected antiproton-to-proton flux ratio at Earth for different dark matter halo profiles and different propagation models, as well as the respective upper bound on the cross-sections from requiring no excess in the PAMELA $\bar p/p$ data at the 95\%C.L. We have also translated the upper bounds on the cross-section into upper bounds on the boost factor in the Milky Way, under the assumption that the Majorana fermion in our model has a thermal relic abundance which coincides with the cosmic dark matter abundance determined by the WMAP.

Given the very special characteristics of this toy model, where the only source of antiproton production is the fragmentation of the weak gauge boson produced in the $2\rightarrow 3$ process, our results can be interpreted as very conservative upper bounds on the cross-sections and boost factors in the Milky Way for more realistic models, where the dark matter particle also couples to quarks and weak gauge bosons. We have illustrated this fact calculating the upper bounds on the boost factor for some concrete benchmark points in the supersymmetric parameter space where the lightest neutralino constitutes the dark matter of the Universe. Our results indicate that in realistic models the constraints on the boost factors are at least one order of magnitude stronger than the constraints derived in this paper for the toy model.

\section*{Acknowledgements}
We thank David Tran for valuable discussions. The work of MG and AI was partially supported by the DFG
cluster of excellence ``Origin and Structure of the Universe.''

\section*{Note added}
During the completion of this paper, a revised version of \cite{Bell:2011eu} appeared, calculating upper bounds
on the $2\to 3$ annihilation cross-section from the PAMELA $\bar p/p$ data in the case $m_{\eta^0}=m_{\eta^\pm}\sim m_{DM}$,
using the corrected results from \cite{Bell:2011if}.

\begin{appendix}

\section{Cross-section for internal electroweak brems\-strahl\-ung}\label{app:sigma}

The to two-to-three annihilation channel $\chi\chi\to W^-\bar e\nu$ can proceed via radiation of a $W$-boson off one of the final state particles, or off the intermediate scalar particle, which yields three diagrams. Since $\chi$ is a Majorana fermion, the two particles in the initial state can be exchanged, so that six diagrams contribute in total, with a relative minus sign for the latter ones. The differential cross-section, multiplied by the relative velocity $v$, is given by
\begin{equation}
 \frac{vd\sigma(\chi\chi\to W^-\bar e\nu)}{dE_WdE_e} = \frac{1}{16(2\pi)^3m_{DM}^2} |\mathcal{M}_1 + \mathcal{M}_2 + \mathcal{M}_3 - \mathcal{M}_4 - \mathcal{M}_5 - \mathcal{M}_6|^2 \;.
\end{equation}
\begin{figure}
 \begin{center}
  \includegraphics{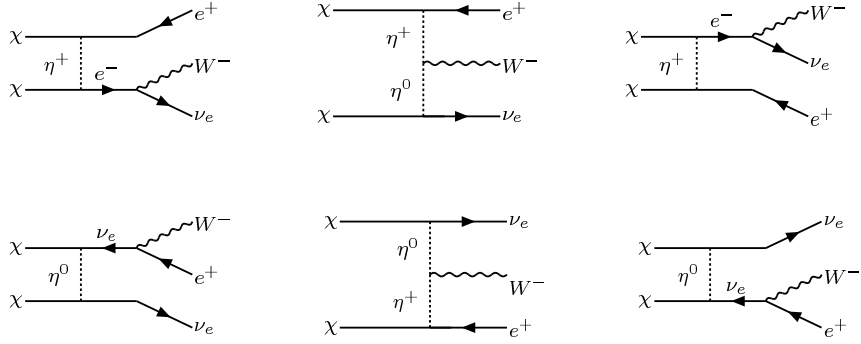}
 \end{center}
 \caption{\label{fig:W_Diagram} Feynman diagrams contributing to the annihilation channel $\chi\chi\to W^-\bar e\nu$.}
\end{figure}
The s-wave contribution can be obtained by setting $p_{\chi_1}=p_{\chi_2}=(m_{DM},0,0,0)$. After expressing the neutrino momentum $p_\nu$ by the other momenta, it is easy to see that all remaining scalar products depend only on the energies $E_W$ of the $W$-boson and $E_e$ of the $\bar e$,  e.g. $p_{\chi_1}\cdot p_W=m_{DM}E_W$. In particular, from the on-shell condition $p_\nu^2=0$ and energy-momentum conservation one obtains
$p_{e}\cdot p_W=m_{DM}(E_e+E_W)-2m_{DM}^2-\frac12 M_W^2$. The $W$-spectrum can be obtained by integrating over the energy $E_e$. The kinematic boundaries are given by $E_e^{min}=m_{DM} - \frac12(E_W + \sqrt{E_W^2 - M_W^2})$ and $E_e^{max}=m_{DM} - \frac12(E_W - \sqrt{E_W^2 - M_W^2})$. The result can be decomposed as
\[ vd\sigma(\chi\chi\to W^-\bar e\nu)/dE_W = vd\sigma_{W_T}/dE_W+vd\sigma_{W_L}/dE_W \,.\]
We find the following result, in terms of $x=E_W/m_{DM}$,
\begin{eqnarray}
\frac{vd\sigma_{W_T}}{dE_W} & = &  \frac{\alpha_{em} f^4}{64 m_{DM}^3\pi ^2 2s_W^2} \left( x_{max}-x \right)\Bigg\{\frac{2\sqrt{x^2-M_W^2/m_{DM}^2}}{(1-x+\mu )^2 F(x,\mu_{\eta^0}) F(x,\mu_{\eta^\pm})} \nonumber\\
& & {} \times \Big[\left(F(x,\mu)+4 \Delta \mu ^2\right) \left(2 (1-x+\mu)^2-F(x,\mu)\right) - 2 \Delta \mu ^2 \left(x^2- {\textstyle\frac{M_W^2}{m_{DM}^2}}\right) \nonumber\\
& & {} - \Delta \mu ^4\Big] - \frac{F(x,\mu)+\Delta \mu ^2}{2(1-x+\mu)^3} \left[ L(x,\mu_{\eta^0}) + L(x,\mu_{\eta^\pm}) \right] \Bigg\} \,, \\
\frac{vd\sigma_{W_L}}{dE_W} & = &  \frac{\alpha_{em} f^4}{64 m_{DM}^3\pi ^2 2s_W^2}\frac{(m_{\eta^0}^2-m_{\eta^\pm}^2)^2}{ 4M_W^2m_{DM}^2 } \Bigg\{ \frac{2 (1-x+\mu )^2-F(x,\mu)-\Delta \mu ^2 }{ 2(1-x+\mu)^3 } \nonumber\\
& & {} \times \left[ L(x,\mu_{\eta^0}) + L(x,\mu_{\eta^\pm}) \right] - 2 \frac{ \sqrt{x^2-M_W^2/m_{DM}^2} }{ (1-x+\mu )^2}\Bigg\} \,,
\end{eqnarray}
where we have defined $\mu_{\eta^\pm}=(m_{\eta^\pm}/m_{DM})^2$, $\mu_{\eta^0}=(m_{\eta^0}/m_{DM})^2$, $\mu=(\mu_{\eta^0}+\mu_{\eta^\pm})/2$, $\Delta \mu=(\mu_{\eta^0}-\mu_{\eta^\pm})/2$, inserted $g=e/s_W$ with $s_W=\sin(\theta_W)$ and $\alpha_{em}=e^2/(4\pi)$, and used
\begin{eqnarray}
  F(x,\mu) & \equiv & (1+\mu )(1+\mu - 2 x)  + M_W^2/m_{DM}^2\,, \nonumber\\
  L(x,\mu) & \equiv & \ln \left[ \frac{ 1-x+\sqrt{x^2-M_W^2/m_{DM}^2} + \mu }{ 1-x-\sqrt{x^2-M_W^2/m_{DM}^2} + \mu} \right]\,.
\end{eqnarray}
The energy of the $W$-boson can vary in the range
\[ M_W/m_{DM} \leq x \equiv E_W/m_{DM} \leq x_{max} \equiv E_W^{max}/m_{DM}=1+M_W^2/(4m_{DM}^2) \,. \]
As expected, the longitudinal contribution is proportional to the square of the $\eta^0\eta^+G^-$-coupling $(m_{\eta^\pm}^2-m_{\eta^0}^2)/M_W$ to the Goldstone mode, and vanishes in the limit $m_{\eta^0}\to m_{\eta^\pm}$. Furthermore, the transversal contribution is proportional to $x_{max}-x$. Thus, the transversal part of the $W$-spectrum is suppressed near the endpoint, while the longitudinal part is not. This behaviour is expected from angular momentum conservation, as was explained in Section \ref{sec:pp}. Note that the charge-conjugated process $\chi\chi\to W^+ e\bar\nu$ produces an identical spectrum of $W^+$ in addition.

The spectrum of $Z$-bosons produced in the electroweak IB annihilation processes $\chi\chi\to Ze\bar e$ and
$\chi\chi\to Z\nu\bar \nu$ can be obtained from the above formulae by replacing $m_{\eta^0}\mapsto m_{\eta^\pm}$ or $m_{\eta^\pm}\mapsto m_{\eta^0}$, respectively, as well as $M_W\mapsto M_Z$. Furthermore, one has to multiply by $2s_W^2\tan^2(2\theta_W)$ for $\chi\chi\to Ze\bar e$ and by $1/(2c_W^2)$ for $\chi\chi\to Z\nu\bar \nu$.

Finally, we note that one can obtain the photon spectrum produced in the electromagnetic IB process $\chi\chi\to \gamma e\bar e$ by replacing $m_{\eta^0}\mapsto m_{\eta^\pm}$, then taking the limit $M_W\to 0$, and multiplying by $2s_W^2$. The resulting formula coincides with Eq.\,9 of Ref.\,\cite{Bringmann:2007nk} (including the prefactor). We have checked that in the limit $m_{\eta^0}\to m_{\eta^\pm}$ our result for the $W$ spectrum is also consistent with the result reported in Ref.\,\cite{Ciafaloni:2011sa} (also including prefactors), and with the result of Ref.\,\cite{Bell:2011if} up to an overall factor of two.

The total cross-sections can be obtained by integrating over the energy of the gauge boson. Here, we just present the result obtained in the limit $m_{\eta^i}\gg m_{DM}$, when keeping $m_{\eta^\pm}^2-m_{\eta^0}^2$ fixed (for our analysis, we use the full expressions),
\begin{eqnarray}
\sigma v(\chi\chi\to W_T^-\bar e\nu) & = &  \frac{\alpha_{em} f^4}{240 m_{DM}^2\pi ^2 2s_W^2 \mu^4} \Big\{(1 - r) \big[1  -14r -94r^2 - 14 r^3 + r^4 \nonumber\\
&&{} + 5 \Delta\mu^2(1+10r+r^2)\big] - 30 r (1+r)(2 r - \Delta\mu^2) \ln(r)\Big\} \,, \\
\sigma v(\chi\chi\to W_L^-\bar e\nu) & = &  \frac{\alpha_{em} f^4}{240 m_{DM}^2\pi ^2 2s_W^2 \mu^4}\frac{5(m_{\eta^0}^2-m_{\eta^\pm}^2)^2}{ 8M_W^2m_{DM}^2 } \Big\{ 1 - 8 r + 8 r^3 - r^4 - 12 r^2 \ln(r) \Big\} \nonumber\,,
\end{eqnarray}
where $r\equiv M_W^2/(4m_{DM}^2)$. The cross-sections for $\chi\chi\to Ze\bar e$ and $\chi\chi\to Z\nu\bar \nu$ can be obtained by the same manipulations as described above for the spectra. In the limit $m_{\eta^i}\gg m_{DM}$, the total cross-section for electromagnetic IB is, for comparison, given by $\sigma v(\chi\chi\to \gamma e\bar e)=\frac{\alpha_{em} f^4}{240 m_{DM}^2\pi ^2 \mu_{\eta^\pm}^4}$.

\end{appendix}

\end{document}